\documentclass[onecolumn,superscriptaddress,aps,prd,nofootinbib,floatfix,10pt]{revtex4-2}

\usepackage[pdftex]{graphicx}
\usepackage[dvipsnames]{xcolor}
\usepackage{color}
\usepackage{amsmath}
\usepackage{mathptmx}
\usepackage{dcolumn}
\usepackage{cancel}
\usepackage{ulem}
\usepackage{bbold}
\usepackage{braket}
\usepackage{lipsum} 
\usepackage{caption}
\usepackage{subcaption}
\usepackage{soul}
\usepackage{CJKutf8}
\usepackage{setspace}
\usepackage{hyperref}
\usepackage{cleveref}
\usepackage[T1]{fontenc}
\definecolor{darkblue}{RGB}{0,0,149}
\hypersetup{
    colorlinks=true,
    linkcolor=red,
    filecolor=darkblue,
    urlcolor=darkblue,
    citecolor=Green,
    linktocpage=true,
}
\usepackage{url}
\usepackage{orcidlink}

\begin{document}
\begin{CJK*}{UTF8}{bsmi}
\title{Acoustic Ginzburg effect with nonclassical states of motion}
 
\author{Hui Wang (王惠)\orcidlink{0000-0003-1012-1124}} 
\email{huiwangph@gmail.com}
\affiliation{Institute for Quantum Science and Engineering, Texas A\&M University, College Station, Texas 77843, USA}

\date{\today}
\begin{abstract}
We explore the acoustic Ginzburg effect using a toy model consisting of a mass-spring chain and a detector. This effect is characterized by the excitation of the detector from its ground state and the generation of acoustic phonons in the chain from the vacuum when the detector travels at a uniform speed greater than the speed of sound. By analyzing a scenario where the detector travels in a superposition of two trajectories at different uniform velocities relative to the center of mass of the chain, we find that the reduced state of either the chain phonons or the detector's internal excitations differ when the detector is in a superposition of distinct trajectories as compared with a single localized trajectory. This distinction enables the chain or the detector to act as a quantum device to distinguish the detector's trajectory.
\end{abstract}

\maketitle
\end{CJK*}

\section{Introduction}

Among the most exciting features of quantum field theory are the zero-point fluctuations of the quantum vacuum, which give rise to the Hawking radiation~\cite{hawking1974black, hawking1975particle}, Unruh effect~\cite{unruh1976notes}, and dynamical Casimir effect~\cite{moore1970quantum, fulling1976radiation}. In addition to these phenomena, Ginzburg radiation, also known as quantum Cherenkov radiation~\cite{ginzburg1945, ginzburg1996radiation,macleod,pieplow2015cherenkov, milton2020self}, predicts that a neutral object moving at superluminal speed relative to a dielectric surface can, under certain conditions, emit photon radiation from the quantum vacuum. Ginzburg radiation is closely related to the quantum friction phenomenon, which involves a neutral particle with an internal degree of freedom that functions as a detector. When this detector, initially in its ground state, moves parallel to a medium surface, it can become excited and emits a surface plasmon or surface phonon with a wave vector in the direction of its motion, leading to a frictional force~\cite{volokitin2007near, pieplow2015cherenkov,  barton2010van, maghrebi2013, milton2020self, lang2022}. 

Relativistic radiation phenomena, such as Ginzburg radiation, the Unruh effect, and the dynamical Casimir effect, have not yet been observed in experiments, partly due to the challenge of inducing a massive object to move at ultra-relativistic speeds. Because of this, analogues of these effects involving acoustic phonons have been studied in various physical systems, including condensed matter systems such as quantum fluids and GaAs nanowires~\cite{marino, tian2021probing, nakayama2023}. These analogues make it feasible for neutral particles to achieve speeds exceeding the acoustic phonon velocity. To investigate acoustic Ginzburg radiation, we employ a mass-spring harmonic oscillator model, which effectively describes an `atomic detector' dipole interacting with a one-dimensional, spatially extended chain of electric dipoles. By taking the continuum limit, the dipole chain is approximated as a $1+1$ dimensional massless scalar field. Under specific conditions, a moving detector would experience quantum friction due to the creation of acoustic phonons in the dipole chain, along with the excitation of the detector's internal degree of freedom. This model enables the exploration of a wide variety of quantum vacuum effects. Specifically, the dynamics of the atomic detector allow for the realization of an acoustic Galilean analogue of Ginzburg radiation, characterized by the creation of acoustic phonons from the quantum vacuum of an elastic medium.


Previous studies have explored the relationship between quantum photon radiation and the detector's nonclassical center-of-mass degrees of freedom for various purposes.
In \cite{biallynicka1991velocity,rzazewski,fedorov}, the effect of the coherent spreading of the center-of-mass wave function of an excited hydrogen-like atom on nonclassical emission spectroscopy was analyzed. In \cite{smith2020quantum,grochowski2021quantum}, an excited atom in a superposition of localized momentum wave packets was used as a clock to demonstrate quantum time dilation in spontaneous emission. Furthermore, \cite{fedorov,sudhir} examined the impact of the dynamic delocalization of an Unruh-DeWitt detector's center-of-mass on photon emission and absorption rates during acceleration. In \cite{barbado,foo2020unruh}, the authors investigated an Unruh-DeWitt detector prepared in a superposition of distinct acceleration trajectories, exploring the coherence of the detector's excited states and interference effects in emission and absorption spectra. However, these papers considered only free space in the Minkowski vacuum, meaning that the detector interacts with a massless scalar field with a continuous photon spectrum. In contrast, our current work adopts a one-dimensional dipole chain interacting with the detector, where the chain is modeled as a $1 + 1$ dimensional massless scalar phonon field, and the phonons have a discrete spectrum due to the finite length of the chain. Furthermore, it is important to note that our scheme operates within the framework of Galilean relativity, making it more feasible to achieve in real experiments. The model proposed in this work is similar to that realized in~\cite{chen2008intrinsic}, where acoustic surface phonons are generated by the interaction of drifting graphene charges with silicon dioxide. In our approach, instead of a charge, a dipole moving close and parallel to the surface of a dielectric solid such as silicon dioxide, modeled here as a dipole chain, will cause a dipole-dipole interaction. This interaction is expected to create a slight distortion of the surface, which will travel with the detector. Assuming negligible thermal noise, we anticipate that surface phonons will be produced from the vacuum in the background of this distortion, leading to acoustic Ginzburg radiation. In addition to a dipole propagating close to and in parallel with a dielectric (insulating) solid surface, experimentally relevant acoustic analogues may be realized with ion-trap experiments as well as other AMO (e.g., BEC) setups in long and narrow trapping potentials~\cite{marino, tian2021probing, andrews, henson, gooding}.

In this paper, by assuming a detector with two internal energy levels and extending the acoustic Ginzburg effect from a detector traveling along a classical trajectory to one traveling in a quantum superposition of two distinct trajectories, we find that the chain field, the internal degrees of freedom of the detector, and the detector's trajectory can be entangled. This is the first investigation of a quantum acoustic counterpart to the Ginzburg effect. Our study is motivated by the question of whether distinguishing between localized and superpositioned particle trajectories requires direct measurement of the particle's motion. We find that observing either the chain phonon excitation or the detector's internal state alone provides an indirect method of distinguishing the detector's trajectory. Additionally, observing the detector's internal state alone provides insights into the state of the chain field. Recent advances in the preparation of coherent momentum superpositions of ytterbium ions~\cite{johnson2017ultrafast}, suggest that realizing superposed trajectories of an atomic detector in experiments is becoming feasible. In a previous study~\cite{wang2024relational}, we utilized the dipole chain as an operational ruler to investigate whether an object is prepared in a localized state or a superposition of different positions. In this work, we aim to develop a systematic approach to explore the behavior of the detector's trajectory. This model also has the potential to provide an operational framework for studying the behavior of quantum fields within quantum reference frames~\cite{giacomini}.

This paper is organized as follows: In Sec.~\ref{chain}, we introduce a toy-model of an atomic detector electromagnetically coupling to a one-dimensional mass-spring chain consisting of $N$ dipoles, and approximate the dipole chain model as a $1 + 1$ dimensional massless scalar field. In Sec.~\ref{classicaltraj}, we concentrate on scenarios where the detector travels with a well-defined velocity. We then explore the classical reaction of the chain and observe the acoustic Ginzburg effect. In Sec.~\ref{supertraj}, we extend our formalism to consider a detector traveling in a quantum superposition of two distinct trajectories. Within this section, Subsec.~\ref{subseca} examines the chain phonon state corresponding to the detector's trajectory, while Subsec.~\ref{subsecb} explores the potential of using the detector's internal state to probe its trajectory. Concluding remarks are provided in Sec.~\ref{conclusion}.

\section{Detector-dipole chain model Hamiltonian}
\label{chain}
\begin{figure}
\begin{center}
\includegraphics[width=\linewidth]{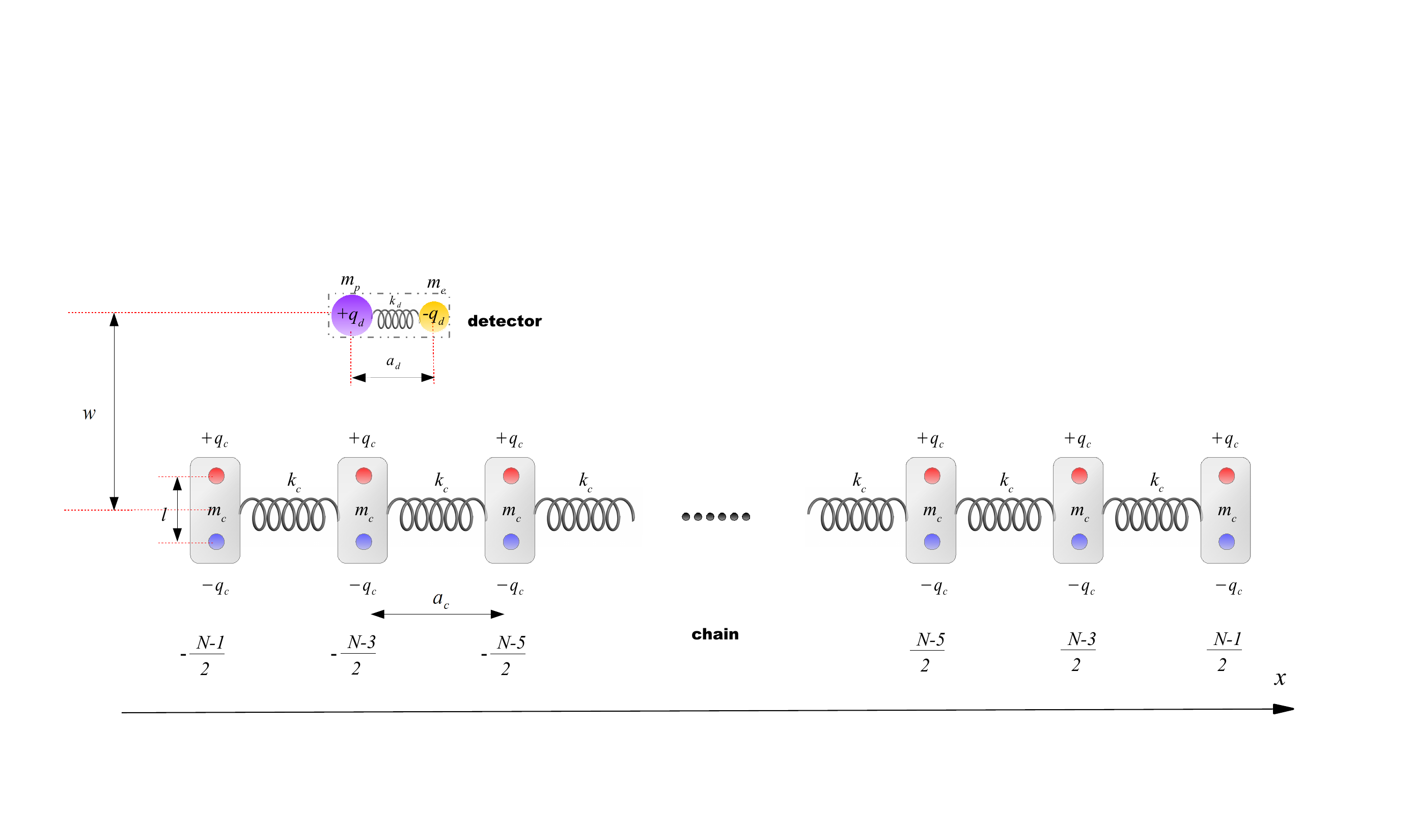} 
\caption{Schematic of the one-dimensional chain and detector system. The chain consists of $N$ electric dipoles with moments $\mathsf{p}_c=q_cl$, with $l$ being the distance between the chain dipole charges $q_c$ and $-q_c$. The dipoles coupled to their nearest neighbors through springs with spring constant $k_c$, and separated at equilibrium by a distance $a_c$. The detector is modeled as a dipole composed of a positive charge $q_d$ with mass $m_p$ and a negative charge (electron) $-q_d$ with a relatively smaller mass $m_e$. The charges are connected by a spring with constant $k_d$ and separated at equilibrium by a distance $a_d$. The detector is positioned a vertical distance $w$ from the chain, and its electrostatic interaction with the chain is short-ranged. Motion is confined to one dimension along the $x$-axis.}
\label{rulerscheme}
\end{center}
\end{figure}

Fig.~\ref{rulerscheme} illustrates the mass-spring harmonic oscillator model, comprising a one-dimensional chain interacting with a detector. The quantities associated with the chain and the detector are denoted with the suffixes `$c$' and `$d$', respectively. The chain consists of $N$ identical dipoles, each with mass $m_c$ and coordinates $x_{c,n}$, where $n = -(N-1)/2, \dots, (N-1)/2$. To ensure symmetry around the chain's center, $N$ is chosen to be an odd number. These dipoles are interconnected by harmonic nearest-neighbor interactions, represented by springs with spring constant $k_c$ and with classical equilibrium separation $a_c$.   

The detector, modeled as a hydrogen-like dipole, consists of a nucleus with a positive charge $q_d$ and mass $m_p$ at position $x_p$, connected via a spring with spring constant $k_d$ to a negative charge (electron) $-q_d$ with relatively smaller mass $m_e$ at position $x_e$. The equilibrium distance between the two charges is $a_d$. The detector and the chain interact through electrostatic forces. This configuration ensures that the interaction between the detector and the chain is of the dipole-dipole type, making it inherently short-ranged. The chain dynamically responds to the detector’s position and motion, and the detector’s internal degree of freedom can become excited through its interaction with the chain.

To derive the detector-chain Hamiltonian in terms of the center-of-mass and relative coordinates, we begin by introducing the chain's center-of-mass coordinate:
\begin{equation}
x_\mathrm{cCM} = \frac{1} {N} \sum_{n=-\frac{N-1}{2}} ^{\frac{N-1}{2}} x_{c,n},
\end{equation}
and the relative displacement of the $n$-th dipole:
\begin{equation}
\phi_{n} = x_{c,n} - x_\mathrm{cCM}-n a_c.
\end{equation}
Here, $x_\mathrm{cCM}$ represents the chain's center-of-mass position, and $\phi_{n}$ gives the displacement of the $n$-th dipole relative to its equilibrium position. The total mass of the chain is $M_c=N m_c$. We simplify the problem by fixing the chain center-of-mass as the observer's reference frame, effectively setting $x_\mathrm{cCM} =0$. This constraint is equivalent to the condition for the relative dipole displacement coordinates: 
\begin{equation}
\sum_{n=-\frac{N-1}{2}}^{\frac{N-1}{2}}\phi_{n}=0.
\label{constrainteq}
\end{equation}

We then describe the detector subsystem using center-of-mass and relative coordinates: 
\begin{align}
x_d=\frac{m_p x_p + m_e x_e} {m_p + m_e},~~ x_{ep}=x_e-x_p.
\end{align}
where $x_d$ is the detector's center-of-mass position, and $x_{ep}$ is the relative position of the two detector components. The total and reduced masses of the detector are $M_d = m_p + m_e$ and $\tilde{m}_d = m_p m_e / (m_p + m_e)$. 

Using these coordinates, we express the total Hamiltonian of the detector-chain system as $\hat{H}_{\mathrm{tot}} =\hat{H}_{\mathrm{det}} +\hat{H}_{\mathrm{chain}} +\hat{V}_{\mathrm{int}}$. The free Hamiltonians for the chain and the detector subsystems are given by:
\begin{equation}
 \hat{H}_\mathrm{chain} = \sum_{n=-\frac{N-1}{2}}^{\frac{N-1}{2}} \frac{\hat{p}_{n}^2}{2m_c} + \frac{1}{2} k_c \sum_{n=-\frac{N-1}{2}}^{\frac{N-3}{2}} (\hat{\phi}_{n+1} - \hat{\phi}_{n})^2 
\label{chainham}   
\end{equation}
and
\begin{equation}
 \hat{H}_\mathrm{det} =  \frac{\hat{p}^2_d}{2 M_d}  +\frac{\hat{p}^2_{ep}}{2\tilde{m}_d} + \frac{1}{2} k_d (\hat{x}_{ep}-a_d)^2, 
\label{detham}   
\end{equation}
These Hamiltonians follow from the Lagrangians derived in Appendix~\ref{coordtrans}, where $\hat{p}_n$, $\hat{p}_d$, and $\hat{p}_{ep}$ are the momentum operators canonically conjugate to the position operators $\hat{\phi}_n$, $\hat{x}_d$, and $\hat{x}_{ep}$, respectively.

We couple the detector to the chain dipoles via dipole-dipole interaction, described by the Coulomb potential energy. The interaction potential is initially presented in its general form and subsequently approximated under the assumption that displacements are small compared to the chain-detector separation. This approximation yields an interaction term dependent on the chain dipole displacements ($\phi_n$) and the detector’s relative coordinate ($x_{ep}$):
\begin{eqnarray}
V_\mathrm{int} &=& \frac{q_d q_c} {4\pi \epsilon_0} \sum_{n=-\frac{N-1}{2}}^{\frac{N-1}{2}} \left[\frac{1} {\sqrt{(w-l/2)^2+(x_p-x_{c,n})^2}} - \frac{1} {\sqrt{(w+l/2)^2+(x_p-x_{c,n})^2}}\right. \cr
&&\left. - \frac{1} {\sqrt{(w-l/2)^2+(x_e-x_{c,n})^2}} + \frac{1} {\sqrt{(w+l/2)^2+(x_e-x_{c,n})^2}} \right] \cr
&\approx& \frac{q_d {\mathsf{p}}_c w}{4\pi \epsilon_0} \sum_{n=-\frac{N-1}{2}} ^{\frac{N-1}{2}} \left[ -\frac{x_{ep}}{a_c} \frac{\partial}{\partial n} (x_n^{\prime 2} + w^2)^{-3/2} + \frac{x_{ep} \phi_n}{a_c^2} \frac{\partial^2}{\partial n^2} (x_n^{\prime 2} + w^2)^{-3/2} \right],
\end{eqnarray}
where $x'_n=x_d - na_c$, and $w$ represents the perpendicular distance between the chain and the detector (for details, see Appendix~\ref{coordtrans}). After the evaluation, the coordinates are quantized to operators to describe the interaction quantum mechanically:
\begin{eqnarray}
\hat{V}_\mathrm{int} \approx \frac{q_d {\mathsf{p}}_c w}{4\pi \epsilon_0} \sum_{n=-\frac{N-1}{2}} ^{\frac{N-1}{2}} \left\{ -\frac{\hat{x}_{ep}}{a_c} \frac{\partial}{\partial n} \left[\left( na_c - \hat{x}_d \right)^2 + w^2\right]^{-3/2} + \frac{\hat{x}_{ep} \hat{\phi}_n} {a_c^2} \frac{\partial^2}{\partial n^2} \left[\left( na_c - \hat{x}_d \right)^2 + w^2\right]^{-3/2} \right\}.
\label{linterac2}
\end{eqnarray}
The approximation is valid under the limit $l, |\phi_n|, |x_{ep}| \ll w$, In this context, the displacements $\phi_n$ are considered small compared to the chain dimensions.

We focus on the regime where the detector-chain separation $w$ is much larger than the chain unit cell length $a_c$, and the chain's classical equilibrium length $L=(N-1)a_c$ is much larger than $w$. These conditions ensure that edge effects are negligible and validate the continuum approximation by allowing the chain to be treated as an infinite medium. To simplify the chain's dynamics, we restrict our analysis to field solutions predominantly composed of the $\alpha$-th modes with wavelengths $\lambda_\alpha=2L/\alpha$ much larger than the dipole spacing $a_c$, such that $\lambda_\alpha \gg a_c$. In this regime, the chain dynamics transition from a discrete description $(\hat{\phi}_n, \hat{p}_n)$ to a continuous field representation $(\hat{\phi}(x,t), \hat{\pi}(x,t))$, with $x = n a_c$ representing the spatial coordinate along the chain. This continuum approximation, described in detail in Appendix~\ref{hamcontlim}, allows the total Hamiltonian of the detector-chain system, originally expressed in discrete variables (Eqs.~(\ref{chainham}), (\ref{detham}), and (\ref{linterac2})), to be reformulated as integrals over continuous fields:
\begin{eqnarray}   
   \hat{H}_{\mathrm{tot}} &=& \frac{1}{2}\int_{-L/2}^{L/2} dx \left( \frac{\hat{\pi}^2} {\rho_c} + \Upsilon_c \nabla\hat{\phi}^2 \right) +\frac{ \hat{p}_d^2} {2M_d} +\frac{\hat{p}_{ep}^2} {2\tilde{m}_d}+ \frac{1}{2} k_d (\hat{x}_{ep}-a_d)^2 \cr
   &&+ g \hat{x}_{ep}\int_{-L/2} ^{L/2}dx\left[ -\frac{\partial}{\partial x} h(x,\hat{x}_d)+\phi  \frac{\partial^2}{\partial x^2}h(x,\hat{x}_d)\right],
\label{continuum1}
\end{eqnarray}
where $\hat{\phi}(x,t) = \hat{\phi}_n/a_c$, $\hat{\pi}(x,t) = \hat{p}_n/a_c$, $\rho_c = m_c/a_c$ representing the chain mass density, and $\Upsilon_c = k_c a_c$ denotes the Young's modulus. We use the shorthand notation $g = {\mathsf{p}}_d {\mathsf{p}}_c w / ({4\pi \epsilon_0 a_d a_c})$ to describe the coupling strength, where ${\mathsf{p}}_d = q_d a_d$ signifies the dipole moment of the detector. To simplify notation, we introduce the function:
\begin{equation}
h(x, \hat{x}_d) = \left[\left( x - \hat{x}_d \right)^2 + w^2\right]^{-3/2}.
\label{hfunction}
\end{equation}
The continuum approximation version of constraint~(\ref{constrainteq}) reads as follows:
\begin{equation} 
\int_{-L/2} ^{L/2}dx\phi(x,t)=0.
\label{constraint1eq}
\end{equation}



\section{Detector with classical trajectory }\label{classicaltraj}

The dynamics of the position operators $\hat{x}_{ep}$, $\hat{x}_d$, $\hat{\phi}$ and the momentum operators $\hat{p}_{ep}$, $\hat{p}_d$, $\hat{\pi}$ are determined by the Heisenberg equations derived from Hamiltonian (\ref{continuum1}), as detailed in Appendix~\ref{semiclameth}. In this section, we explore the dynamics of these operators under the assumption that the detector is point-like and travels along a single localized trajectory.

We begin by applying the semiclassical-like approximation, decomposing the detector's internal degrees of freedom ($\hat{x}_{ep}$, $\hat{p}_{ep}$) and the chain field operators ($\hat{\phi}$, $\hat{\pi}$) into their classical mean coordinates ($\bar {x}_{ep}$, $\bar {p}_{ep}$, $\bar {\phi}$, $\bar {\pi}$) and quantum fluctuation components ($\delta \hat{x}_{ep}$, $\delta \hat{p}_{ep}$, $\delta \hat{\phi}$, $\delta \hat{\pi}$):
\begin{eqnarray}
\hat{x}_{ep} &=& \bar {x}_{ep}+\delta \hat{x}_{ep}\\
\hat{p}_{ep} &=& \bar {p}_{ep}+\delta \hat{p}_{ep}\\
\hat{\phi} &=& \bar {\phi}+\delta \hat{\phi}\\
\hat{\pi} &=& \bar {\pi}+\delta \hat{\pi}.
\label{decomposition}
\end{eqnarray}
Under the condition that the detector's trajectory is localized, its center-of-mass coordinates $\hat{x}_d$ and $\hat{p}_d$ can be treated classically, i.e., $\hat{x}_d=\bar{x}_d$ and $\hat{p}_d=\bar{p}_d$. 

We then decompose the Heisenberg equations involving the operators ($\hat{x}_{ep}$, $\hat{p}_{ep}$, $\hat{\phi}$, $\hat{\pi}$, $\hat{x}_d$, $\hat{p}_d$) into two sets: one for the mean coordinates of the detector and field, and another describing the quantum fluctuations, which depend on these mean coordinates. Detailed derivations of these equations are provided in Appendix~\ref{semiclameth}. In Sec.~\ref{subsec1}, we analyze the solutions for the mean coordinates to explore the classical reaction of the dipole chain. In Sec.~\ref{subsec2}, we derive the Hamiltonian accurate to second order in quantum fluctuation terms, based on the Heisenberg equation for the quantum fluctuation coordinates, and discuss the resulting emergence of acoustic Ginzburg radiation.

\subsection{Classical reaction of the chain}\label{subsec1}

Assuming a large detector mass $M_d$, a stiff detector spring constant $k_d$, and weak field-detector coupling $g$, the effect of the chain field-detector interaction on the mean dynamics of the detector's center-of-mass and its internal degree of freedom can be neglected. The mean coordinate solutions for the detector's internal degree of freedom, following the Heisenberg equations~(\ref{avephieq1}), are as follows:
\begin{eqnarray}
    \bar{p}_{ep}(t) &=& 0, \cr
    \bar{x}_{ep}(t) &=& a_d.
\label{semiheisenberg}
\end{eqnarray}
Defining the detector's center-of-mass as initially positioned at $x_0$ and moving with a constant velocity $v$. The corresponding classical momentum and position of the detector are given by:
\begin{eqnarray}
\bar{p}_d(t) &=& M_d v, \cr \bar{x}_d (t) &=& x_0+ v t.
\end{eqnarray}

\begin{figure}[htbp]
    \centering
    \begin{subfigure}{0.48\linewidth}
        \centering
        \includegraphics[width=\linewidth]{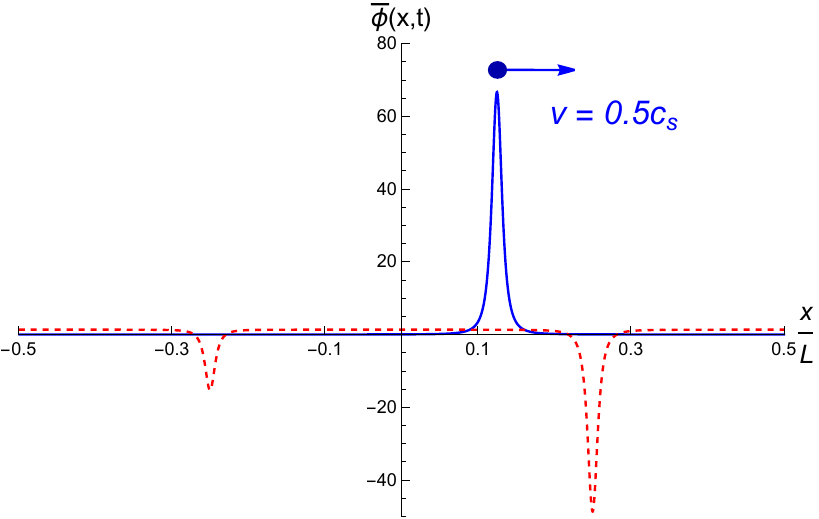}
        \caption{}
        \label{averagephi1}
    \end{subfigure}
    \hfill
    \begin{subfigure}{0.48\linewidth}
        \centering
        \includegraphics[width=\linewidth]{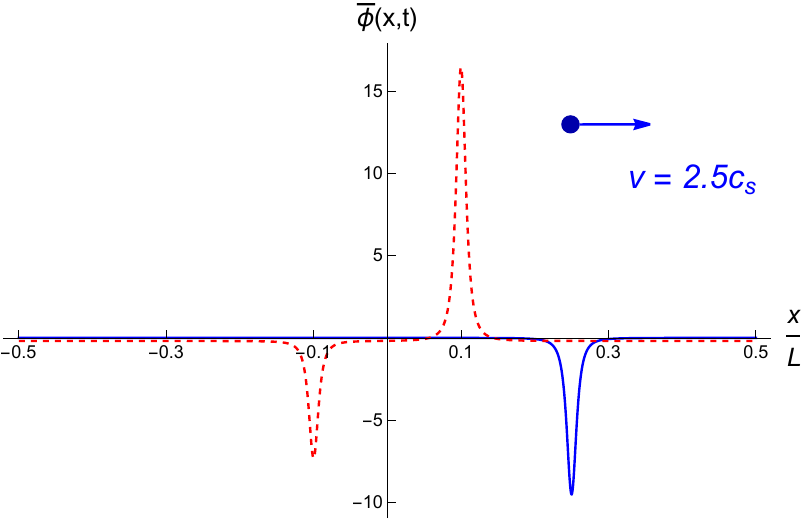}
        \caption{}
        \label{averagephi2}
    \end{subfigure}
    \caption{Plots of the average, longitudinal dipole displacement coordinate $\bar{\phi}(x,t)$ (arbitrary units) vs $x$ (in units of chain length $L$). The detector-chain interaction is turned on at $t=0$ (in units $L/c_s$). At this moment, the detector is located at $x_0=0$, and the perpendicular distance between the detector and the chain is $w=0.01 L$. (a) The case at $t=0.25$ and the detector, which is represented by the blue dot, moving at a speed of $v=0.5c_s$. (b) The case at $t=0.1$ and the detector moving at a speed of $v=2.5c_s$. The blue solid peak represents the first term of Eq.~(\ref{phifinal1}), whose peak coincides with the center-of-mass of the detector. The red dashed peaks represent the second and third terms of Eq.~(\ref{phifinal1}). It is worth noting that the integration of these three terms sums to zero at any time point $t$, as required by constraint~(\ref{constraint1eq}).}
    \label{fig:averagephi}
\end{figure}

Supposing that the chain-detector interaction starts at $t=0$, and that the chain is initially in its undeformed (vacuum) state, with the detector's internal degree of freedom initially in the ground state. Under these conditions, the mean value of the chain field takes the form (see Appendix~\ref{semiclameth} for the derivation details):
\begin{equation}
    \bar{\phi}(x,t)=-g \sum_{\alpha=1}^{N-1} u_{\alpha}(x)\int_0^{t} dt' \frac{\sin \left[\Omega_{\alpha} (t-t')\right]}{\Omega_{\alpha}} \int_{-L/2}^{L/2} dx' \frac{\partial^2}{\partial x^{\prime 2}} h[x',\bar{x}_d(t')]u_{\alpha}(x'),
\label{avephieq}
\end{equation}
where the function $h$ is introduced in Eq.~(\ref{hfunction}), and we use $h[x',\bar{x}_d(t')]$ instead of $h[x',\hat{x}_d(t')]$ because the operators are treated classically. Additionally, the parameters are defined as follows:
\begin{eqnarray}
    &&\Omega_\alpha=2\sqrt{\frac{k_c}{m_c}} \sin \left(\frac{\alpha\pi a_c}{2L}\right),\cr
    &&u_\alpha(x)=\sqrt{\frac{2}{L}}\cos\left[\frac{\alpha\pi }{L} \left(x+\frac{L}{2}\right)\right],\cr
    &&\alpha=1,2,\dots, N-1.
\label{rulermodeseq1}
\end{eqnarray}
Here, for each mode $\alpha$, $\Omega_\alpha$ represents the normal acoustic mode frequency of the chain, and $u_{\alpha}(x)$ is the corresponding orthonormal eigenfunction of the chain's mode in the continuum approximation. Any phonon mode $\alpha$ with wavelength $\lambda_\alpha$ that is smaller than $w$ is effectively filtered out because the high-frequency modes decay before reaching and interacting with the detector. Considering the restrictions in the last paragraph of Sec.~\ref{chain}, the condition $\lambda_\alpha\geq w \gg a_c$ is therefore satisfied for the acoustic phonon modes considered. In this regime, where $\alpha\pi a_c/2L\ll 1$, the sine term can be approximated as $\sin \left(\alpha\pi a_c/2L\right)\approx\alpha\pi a_c/2L$. Consequently, $\Omega_\alpha \approx \alpha\pi c_s/L$ with $c_s= a_c\sqrt{k_c/ m_c}$ is the longitudinal sound speed for chain phonons, and $u_{\alpha}(x)$ can be expressed in terms of $\Omega_\alpha$ approximately as:
\begin{eqnarray}
    u_{\alpha}(x) &\approx& \sqrt{\frac{2}{L}}\cos\left[\frac{\Omega_\alpha}{c_s} \left(x+\frac{L}{2}\right)\right].
    \label{rulermodeseq2}
\end{eqnarray}
 Assuming the detector is well away from the chain edges (i.e., $|\bar{x}_d(t)| \ll L$) and $w \ll L$, through a series of calculations and approximations (detailed in Appendix~\ref{localizeatom}), the simplified analytical solution of $\bar{\phi}(x,t)$ is: 
\begin{eqnarray}
     \bar{\phi}(x,t)=\frac{g a_d} {\rho_c} \left[\frac{h(x,\bar{x}_d)} {c_s^2-v^2} -\frac{h(x,x_0+c_st)}{2c_s (c_s-v)}  -\frac{h(x,x_0-c_st)}{2c_s (c_s+v)} \right].
\label{phifinal1}
\end{eqnarray}
Figure~\ref{fig:averagephi} shows plots of $\bar{\phi}(x,t)$ versus $x$, with $x$ in units of the chain length $L$ and $t$ in units of $L/c_s$. In Fig.  \ref{averagephi1}, $\bar{\phi}(x,t)$ is shown for a detector moving at a speed of $v=0.5c_s$ at $t=0.25$, while Fig.  \ref{averagephi2} presents the results for $v=2.5c_s$ at $t=0.1$.

The three terms enclosed in the parentheses of Eq.~(\ref{phifinal1}) indicate the existence of three traveling wave packets in the chain. The first term corresponds to a wave packet traces out the trajectory $\bar{x}_d(t)=x_0+vt$ of the detector. This is depicted by the blue wave packet, which travels in alignment with the detector's position (the blue dot). The second and third terms enclosed in the curly brackets of Eq.~(\ref{phifinal1}) describe two wave packets that originate from $x_0=0$ at the initial time $t=0$. The wave packet corresponding to the second term travels to the right at the speed of sound, $c_s$ (illustrated by the red peak in the right-half plane), while the wave packet corresponding to the third term travels to the left at the same speed (depicted by the red peak in the left-half plane). These red peaks represent the sound wave `ripples' resulting from the sudden switch-on of the detector-chain interaction. The width of each wave packet is $w$,  corresponds to the spatial extent of the function $h$ and is equal to the vertical distance between the detector and the chain. 
We note that the appearance of these wave packets is not related to quantum friction; instead, it is purely a classical effect. Furthermore, the detector’s momentum remains unchanged since the net displacement of the chain's dipoles integrates to zero; see Eq.~(\ref{constraint1eq}).


\subsection{Quantum Dynamics of the Chain-Detector System: Acoustic Ginzburg Effect}\label{subsec2}

Let us now take a look at the quantum dynamics of the chain-detector system. The Heisenberg equations for the position operators of the detector's internal degree of freedom and the quantum field, $\delta \hat{x}_{ep}$ and $\delta\hat{\phi}$, as well as the corresponding quantum momentum operators, $\delta \hat{p}_{ep}$ and $\delta\hat{\pi}$ (see Appendix~\ref{semiclameth} for details), can be derived from an approximate Hamiltonian that is second-order in quantum fluctuations. This `second-order quantum fluctuation Hamiltonian' is formulated as a sum of terms quadratic in the quantum fluctuations:
\begin{align}    
    \hat{H}_\mathrm{Q} = \frac{1}{2}\int_{-L/2}^{L/2} dx \left[\frac{\delta\hat{\pi}^2} {\rho_c} + \Upsilon_c (\nabla\delta \hat{\phi})^2 \right] +\frac{\delta \hat{p}_{ep}^2}{2\tilde{m}_d} + \frac{1}{2} k_d \delta \hat{x}_{ep}^2 + g \int_{-L/2}^{L/2} dx \delta\hat{\phi} \delta \hat{x}_{ep}\frac{\partial^2 }{\partial x^2} h(x,\bar{x}_d),
    \label{hamiq1}
\end{align}
In this Hamiltonian, the first term represents the free quantum chain field, while the second and third terms account for the free internal degrees of freedom of the detector. The last term describes the interaction between the quantized chain field and the detector. Certain terms, including $\bar{\phi}(x,t)$ and $\partial^3 h(x,\bar{x}_d) /\partial x^3$, are omitted because both functions are even with respect to $\bar{x}_d$ under the condition $L\gg w, |\bar{x}_d|$.

We next proceed to the second quantization of the fluctuation components of the chain field, expressing them in terms of the creation and annihilation operators $\hat{a}_{\alpha}^{\dag},~\hat{a}_{\alpha}$ for the $\alpha$-th phonon mode:
\begin{equation}
\delta\hat{\phi}=\sum_{\alpha=1}^{N-1}\sqrt{\frac{\hbar}{2\rho_c\Omega_{\alpha}}} \left(\hat{a}_{\alpha}+ \hat{a}_{\alpha}^{\dag} \right) u_{\alpha}(x),
\label{quantiz1}
\end{equation}
\begin{eqnarray}
\delta\hat{\pi}&=&-i \sum_{\alpha=1} ^{N-1} \sqrt{\frac{\hbar \rho_c\Omega_{\alpha}}{2}} \left(\hat{a}_{\alpha} -\hat{a}_{\alpha}^{ \dag} \right) u_{\alpha}(x).
\label{quantiz2}
\end{eqnarray}
Furthermore, the fluctuation part of the detector's internal degree of freedom can be modeled as a quantum oscillator with frequency $\omega_d$, expressed in terms of the creation/annihilation operators $b^\dag,~b$:
\begin{equation}
\delta \hat{x}_{ep}=\sqrt{\frac{\hbar}{2\tilde{m}_d\omega_d}} \left(\hat{b}+ \hat{b}^{\dag} \right),
\label{quantiz3}
\end{equation}
\begin{equation}
\delta \hat{p}_{ep}=-i\sqrt{\frac{\hbar\tilde{m}_d\omega_d}{2}} \left(\hat{b}- \hat{b}^{\dag} \right).
\label{quantiz4}
\end{equation}
The total Hamiltonian in Eq.~(\ref{hamiq1}) is given by $\hat{H}_Q=\hat{H}_0+\hat{H}_I$, where $\hat{H}_0$ describes the independent evolution of the chain field and the detector, and $\hat{H}_I$ accounts for their interaction. Substituting Eqs.~(\ref{quantiz1}-\ref{quantiz4}) into $\hat{H}_0$, the free Hamiltonian is quantized as:
\begin{equation}    
 \hat{H}_0 =\sum_{\alpha=1}^{N-1} \hbar\Omega_{\alpha} \hat{a}_{\alpha}^{\dag} \hat{a}_{\alpha} + \hbar\omega_d \hat{b}^{\dag} \hat{b} .
\label{freehami2}
\end{equation}
We next quantize the interaction term from Eq.~(\ref{hamiq1}) and transform it into the interaction picture, yielding the interaction Hamiltonian:
\begin{align}    
    \hat{H}_\mathrm{I}^\mathrm{int} = g \int_{-L/2}^{L/2} dx \frac{\partial^2}{\partial x^2} h(x,\bar{x}_d) \sum_{\alpha=1}^{N-1} \frac{\hbar} {2}\sqrt{\frac{1} {\rho_c \hbar \Omega_{\alpha} \omega_d \tilde{m}_d}}(\hat{a}_{\alpha}e^{-i\Omega_{\alpha} t} + \hat{a}_{\alpha}^{\dag} e^{i\Omega_{\alpha} t}) (\hat{b} e^{-i\omega_d t} + \hat{b}^{\dag} e^{i\omega_d t})u_{\alpha}(x).
\label{vinteraction}
\end{align}
Integrating by parts and using the fact that $\left.\partial^n h(x,\bar{x}_d)/\partial x^n\right|_{x=\pm L/2}\approx 0$ for cases where $w\ll L$, along with the additional assumption that the detector is significantly distant from the chain edges (i.e., $|\bar{x}_d(t)|\ll L$), the lower and upper integration limits in terms of $x$ can be approximated as $\mp \infty$. Consequently, the Hamiltonian is simplified to:
\begin{align}     
\hat{H}_\mathrm{I}^\mathrm{int}=\sum_{\alpha=1}^{N-1} g_\alpha (\hat{a}_{\alpha}e^{-i\Omega_{\alpha} t} + \hat{a}_{\alpha}^{\dag} e^{i\Omega_{\alpha} t}) (\hat{b}e^{-i\omega_d t} + \hat{b}^{\dag} e^{i\omega_d t}) \cos \left[\frac{\Omega_\alpha}{c_s} \left(\bar{x}_d(t)+\frac{L}{2}\right)\right],
    \label{vinteraction2}
\end{align}
with $\bar{x}_d(t)=x_0+vt$. The renormalized ion-ruler mode coupling strength is defined as
\begin{equation} 
g_\alpha=-\frac{g\hbar \Omega_\alpha} {w^2c_s^2} \sqrt{\frac{2\Omega_{\alpha}} {\rho_c L \tilde{m}_d\omega_d}}f(\Omega_{\alpha} w/c_s),
\end{equation}
where $f(y)=y K_1(y)$ is a cut-off function at around $y\sim 1$, with $f(0)=1$ and where  $K_1(y)$ is the modified Bessel function of the second kind:
\begin{equation}
 K_1(y)=y \int_0^{\infty} dx (x^2+y^2)^{-3/2} \cos x.
\label{bessel1eq}
\end{equation}
The cut-off function $f(\Omega_{\alpha} w/c_s)$ suppresses phonon modes with wavelengths smaller than $w$, effectively filtering out these shorter wavelengths. 

Recall that in the dipole chain, the acoustic phonons travel at the sound speed $c_s$, which is analogous to the speed of light in the original Ginzburg effect; it is in principle possible for the detector to move at speeds $|v|>c_s$, resulting in acoustic Ginzburg radiation~\cite{ginzburg1945}. We again assume that the detector is moving at a localized velocity $v$ in the chain's reference frame. In the regime $|v|>c_s$, we assume that the following resonance condition is satisfied for the $\alpha_0$-th phonon mode:
\begin{eqnarray}  
\frac{v\Omega_{\alpha_0}}{c_s}=\Omega_{\alpha_0}+\omega_d.
\label{resonance0}
\end{eqnarray}  
We then apply the rotating wave approximation (RWA), which is valid in the weak-coupling regime. Due to the sudden switch-on, nonzero phonon mode excitations could occur, challenging the validity of the RWA. To address this, a possible experimental approach is to first position the atomic detector near a ‘cold’ surface, allowing it to equilibrate with the surface, which remains in a shifted vacuum state due to the atom’s presence. By gradually accelerating the detector from rest using a smooth switching function, the system could reach a uniform speed, thereby better adhering to the approximation and justifying the RWA. Given that we are considering finite $N$, we can assume a discrete spectrum for the chain phonons. This allows the interaction terms between the detector and phonon modes $\alpha \neq \alpha_0$ to be discarded, and the interaction Hamiltonian  (\ref{vinteraction2}) simplifies to
\begin{equation}    
    \hat{H}_\mathrm{I,\alpha_0}^\mathrm{int} =\frac{g_{\alpha_0}}{2} (\hat{a}_{\alpha_0}\hat{b} + \hat{a}_{\alpha_0}^{\dag}  \hat{b}^{\dag}).
    \label{ndpaham}
\end{equation}

The reduced Hamiltonian~(\ref{ndpaham}) coincides with that of a non-degenerate parametric amplifier (NDPA). When the chain and detector system begins in the vacuum state, two simultaneous events occur during the NDPA process: the detector absorbs a photon and transitions to an excited state, while a corresponding phonon is generated within the dipole chain, extracting energy from the detector's kinetic motion. From the above, we can see that when the detector moves at a speed exceeding the acoustic phonon velocity of the chain, the interaction Hamiltonian can be effectively described by Eq.~(\ref{ndpaham}), enabling the generation of photon-phonon pairs from vacuum. Consequently, this results in the emergence of acoustic Ginzburg radiation.

\section{Detector with superposition of trajectories}
\label{supertraj}
In this section, we extend our discussion from a detector following a single localized trajectory to scenarios where the detector is prepared in a quantum superposition of localized trajectories with different velocities. To ensure the validity of the RWA, we neglect the momentum uncertainty of each detector trajectory in this analysis. We examine the reduced states of the chain phonons and the detector's internal vibrational degrees of freedom, either of which can be excited during the Ginzburg effect. These excitations can act as indirect measurements of whether the detector follows a classical (localized) trajectory or a quantum trajectory (a superposition of two localized paths).

We will focus on the following scenario: the detector is initially in a superposition of position states with coordinates $x_{01}$ and $x_{02}$ at time $t=0$. These positions correspond to two components of the detector traveling with different velocities, $v_1$ and $v_2$, respectively. The momenta of these components, $p_{d_{1}} = M_d v_1$ and $p_{d_{2}} = M_d v_2$ are determined by their respective velocities. While the Ginzburg radiation extracts energy from the detector's kinetic motion, we assume for simplicity that the back-action on the detector's center-of-mass dynamics due to the chain field-detector interaction is negligible in our analysis. Additionally, we assume that the uncertainty in the detector's individual trajectory state is smaller than the dimensions of the chain, allowing us to treat the detector’s superposition of two trajectories as well-defined. To preserve the classical nature of the chain's average response to the detector, we additionally assume that the positional difference between the two detector trajectories remains less than the widths of the "ripples" depicted in Fig.~\ref{fig:averagephi}. Specifically, this assumption requires $|x_{01}-x_{02}+(v_1-v_2)t|<w$. 

When the detector travels along quantum trajectories, its coordinates must be treated as quantum operators. The approximate Heisenberg equations of motion for $\hat{x}_d$ and $\hat{p}_d$ are as follows:
\begin{eqnarray}
    \dot{\hat{p}}_d(t) &=& 0,\cr
   \dot{\hat{x}}_d(t)&=& \frac{\hat{p}_d(0)}{{M}_d}.
\end{eqnarray}
These equations indicate that the detector's kinetic energy term must be included in the second-order quantum fluctuation Hamiltonian~(\ref{hamiq1}). Additionally, in Hamiltonian (\ref{hamiq1}), $h(x, \bar{x}_d)$ should be replaced by $h(x, \hat{x}_d) = \left[ \left( x - \hat{x}_d \right)^2 + w^2 \right]^{-3/2}$, where $\hat{x}_d$ is a quantum operator. In Sec.~\ref{subseca}, we demonstrate that measuring the chain phonon state allows one to distinguish between a localized trajectory and a superposed trajectory of the detector. However, a detector with a single internal degree of freedom cannot achieve this distinction. In Sec.~\ref{subsecb}, we consider a detector with two internal degrees of freedom. Under certain resonance conditions, this detector can simultaneously monitor both its trajectory and the chain phonon state.

\subsection{Chain-Based Measurement of the Detector's Trajectory}\label{subseca}
By adding the term including the quantum operators $\hat{p}_d$ back to Hamiltonian~(\ref{hamiq1}), replacing $h(x, \bar{x}_d)$ with $h(x, \hat{x}_d) = \left[ \left( x - \hat{x}_d \right)^2 + w^2 \right]^{-3/2}$, and assuming the detector is significantly distant from the chain edges, the resulting quantum Hamiltonian, expressed in terms of creation and annihilation operators, can be approximated as:
\begin{align}    
     \hat{H}_\mathrm{Q} = \frac{\hat{p}_d^2} {2M_d}+ \sum_{\alpha=1}^{N-1} \hbar\Omega_{\alpha} \hat{a}_{\alpha}^{\dag}\hat{a}_{\alpha} + \hbar\omega_d \hat{b}^{\dag}\hat{b}+\sum_{\alpha=1}^{N-1} g_\alpha (\hat{a}_{\alpha} + \hat{a}_{\alpha}^{\dag} ) (\hat{b} + \hat{b}^{\dag}) \cos \left[\frac{\Omega_\alpha}{c_s} \left(\hat{x}_d+\frac{L}{2}\right)\right].
    \label{hamiq2}
\end{align}
The last term in Eq.~(\ref{hamiq2}) entangles the quantum field (represented by $\hat{a}_{\alpha}^{\dag}, \hat{a}_{\alpha}$), the detector's internal degree of freedom (represented by $\hat{b}^{\dag}, \hat{b}$), and the detector's position $\hat{x}_d$. The remaining terms in Eq.~(\ref{hamiq2}) constitute the free Hamiltonian. In this case, the free Hamiltonian differs from Eq.~(\ref{freehami2}) and is expressed as:
\begin{equation}    
 \hat{H}_0 =\frac{\hat{p}_d^2} {2M_d} + \sum_{\alpha=1}^{N-1} \hbar\Omega_{\alpha} \hat{a}_{\alpha}^{\dag} \hat{a}_{\alpha} + \hbar\omega_d \hat{b}^{\dag} \hat{b} .
\label{freehami3}
\end{equation}
The initial state of the chain-detector system is given by:
\begin{align}
|\psi^s(0)\rangle= \frac{1}{\sqrt{2}} \int_{-L/2}^{L/2}dx \left[\delta\left(\hat{x}-x_{01}\right)+ \delta\left(\hat{x}-x_{02}\right) \right] |x\rangle_d \otimes|g\rangle \otimes \prod_{\alpha=1}^{N-1} |0\rangle_{\alpha}.
\label{inistate}
\end{align}
Here, $|x\rangle_d$ represents the position state of the detector. Under the free-particle time evolution operator, it evolves as $\exp(-{i\hat{p}_d^2t}/{2M_d}) |x\rangle_d= |x+{p_d t} /{M_d}\rangle_d$. $|0\rangle_{\alpha}$ is the vacuum state of the $\alpha$-th phonon field mode, and $|g\rangle$ represents the ground state of the detector's single internal excitation frequency $\omega_d$. We assume the detector to be point-like and initially in an equally weighted coherent superposition of two position states. A more general and realistic case, where the detector is modeled as a superposition of two detector trajectories with a relative phase and different weights, is discussed in Appendix~\ref{systate}.

Furthermore, we consider the regime where $v_2>v_1>c_s$ and the resonance conditions
\begin{eqnarray}  
\frac{v_1\Omega_{\alpha_1}}{c_s}&=&\Omega_{\alpha_1}+\omega_d,\cr
\frac{v_2\Omega_{\alpha_2}}{c_s}&=&\Omega_{\alpha_2}+\omega_d
\label{resonance}
\end{eqnarray}  
are satisfied for the $\alpha_1$th and $\alpha_2$th field mode. 

In the weak coupling limit, where $g_{\alpha_1}$, $g_{\alpha_2}\ll 1$, perturbation theory can be applied. Accordingly, the chain-detector system in the Schrödinger picture at time $t$, to first order, is described approximately by the following state. Calculation details are provided in Appendix~\ref{subsec:single}:
\begin{eqnarray}
|\psi^s(t)\rangle&\approx&\frac{1}{\sqrt{2}}|x_{01}+v_1t\rangle_d \otimes \left( |g\rangle\otimes \prod_{\alpha=1}^{N-1}|0\rangle_{\alpha} -\frac{ig_{\alpha_1}t }{2} |e\rangle \otimes\prod_{\alpha\neq\alpha_1} |0\rangle_{\alpha} |1\rangle_{\alpha_1} \right)\cr
&&+ \frac{1}{\sqrt{2}} |x_{02} + v_2t\rangle_d \otimes \left(|g\rangle \otimes \prod_{\alpha=1}^{N-1} |0\rangle_{\alpha} -\frac{ig_{\alpha_2}t }{2} |e\rangle \otimes \prod_{\alpha\neq\alpha_2} |0\rangle_{\alpha} |1\rangle_{\alpha_2} \right).
\label{superpos}
\end{eqnarray}
From Eq.~(\ref{superpos}), it becomes clear that the detector's trajectory (velocity) becomes entangled with the frequency of the excited chain phonon. By tracing out the detector's position state and internal degrees of freedom, and selectively tracing out the chain phonon modes except for $\alpha=\alpha_1, \alpha_2$, the reduced state for the phonon modes $\alpha_1$ and $\alpha_2$ is given by
\begin{eqnarray}
\rho_c^s(t)\approx \frac{1}{\mathcal{N}} \left\{|0\rangle_{\alpha_1} \langle 0|\otimes |0\rangle_{\alpha_2} \langle 0|+ \frac{1}{8} \left[(g_{\alpha_1} t)^2|1\rangle_{\alpha_1} \langle 1| \otimes |0\rangle_{\alpha_2} \langle 0|+ (g_{\alpha_2} t)^2|0\rangle_{\alpha_1} \langle 0|\otimes |1\rangle_{\alpha_2} \langle 1|\right] \right\},
\label{redsuperpos}
\end{eqnarray}
where $\mathcal{N}=1+[(g _{\alpha_1} t)^2+(g _{\alpha_2} t)^2]/8$. It is evident that both the $\alpha_1$th and $\alpha_2$th phonon modes have a nonzero transition rate to the excited state for the superposed trajectory. In contrast, when the detector follows a localized trajectory, it is not possible for both phonon modes to be excited. For example, when the trajectory has velocity $v_1$, the reduced state for the phonon modes $\alpha_1$ and $\alpha_2$ would be:
\begin{eqnarray}
\rho_c^l(t) \approx \frac{1}{1+(g_{\alpha_1} t)^2/4} \left[|0\rangle_{\alpha_1} \langle 0|+ \frac{1}{4} (g_{\alpha_1} t)^2|1\rangle_{\alpha_1} \langle 1|\right]\otimes |0\rangle_{\alpha_2} \langle 0|.
\end{eqnarray}
Obviously, in this case, only the $\alpha_1$th phonon mode has a nonzero transition rate, while the $\alpha_2$th phonon mode would always remain in its ground state. Therefore, the chain phonon state can function as a measurement device that distinguishes a localized trajectory of the detector from a superposed trajectory of the detector. 

It is natural to ask whether the detector's internal state can also function as a measurement device for the detector's trajectory. In the case where the detector moves with a superposition of velocities $v_1$ an $v_2$, by tracing out the detector's position state and all the chain phonon modes, we obtain the reduced density matrix describing the detector's internal degree of freedom:
\begin{eqnarray}
\rho_d^s(t) \approx \frac{1}{\mathcal{N}} \left\{|g\rangle \langle g|+ \frac{1}{8}  \left[(g _{\alpha_1} t)^2 + (g _{\alpha_2} t)^2 \right] |e\rangle \langle e|\right\}.
\label{reduced0}
\end{eqnarray}
In the case where the detector travels at a localized trajectory with velocity $v_1$, the density matrix of the detector's reduced state would be 
\begin{eqnarray}
\rho_d^l(t) \approx \frac{1}{1+(g_{\alpha_1} t)^2/4} \left[|g\rangle \langle g|+ \frac{1}{4} (g _{\alpha_1} t)^2 |e\rangle \langle e|\right].
\label{reducedloc}
\end{eqnarray}
The detector's internal state would always be a mixture of the ground state and the excited state for both cases. The only difference observed in the reduced state of the detector, whether traveling along a classical trajectory or a superposed trajectory, would be the excitation probability of the detector. However, this probability difference is expected to be small, making the detector's internal state a poor candidate for a reliable measurable distinction between the two scenarios.

\subsection{Detector-Based Measurement of Detector's Trajectory Using Two Internal Excitation Levels}\label{subsecb}
In the previous subsection, we showed that the chain phonon state contains information about the detector's trajectory. In this subsection, we explore the possibility of using the detector's internal state to monitor both the chain's phonon state and the detector's trajectory as it undergoes quantum dynamics.

To obtain a more direct signature of a superposition of trajectories with the detector's internal state, we assume the detector possesses two internal excitation frequencies, $\omega_{d_1}$ and $\omega_{d_2}$. The Hamiltonian Eq.~(\ref{hamiq2}) is then modified as follows:
\begin{eqnarray}    
     \hat{H}_\mathrm{Q} = \frac{\hat{p}_d^2} {2M_d}+ \sum_{\alpha=1}^{N-1} \hbar\Omega_{\alpha} \hat{a}_{\alpha}^{\dag}\hat{a}_{\alpha} + \sum_{i=1,2} \hbar\omega_{d_i} \hat{b}_i^{\dag} \hat{b}_i+ \sum_{i=1,2} \sum_{\alpha=1}^{N-1} g_{\alpha, i} (\hat{a}_{\alpha} + \hat{a}_{\alpha}^{\dag} ) (\hat{b}_i + \hat{b}_i^{\dag})\cos \left[\frac{\Omega_\alpha}{c_s} \left(\hat{x}_d+\frac{L}{2}\right)\right].
\end{eqnarray}
Assuming that the detector travels at a superposition state of $v_1$ and $v_2$. The initial state of the chain-detector system is:
\begin{eqnarray}
|\psi^s(0)\rangle = \frac{1}{\sqrt{2}} \int_{-L/2}^{L/2}dx \left[\delta\left( \hat{x}-x_{01}\right)+ \delta\left(\hat{x}-x_{02}\right)\right]|x\rangle_d \otimes|gg\rangle \otimes \prod_{\alpha=1}^{N-1} |0\rangle_{\alpha},
\end{eqnarray}
where $|gg\rangle$ denotes the ground state of the detector's two internal excitation levels, $\omega_{d_1}$ and $\omega_{d_2}$. The detailed calculation for the case in which the detector is described as a superposition of two trajectories, including a relative phase and differing weights, is provided in Appendix~\ref{subsec:two}.

The resonance conditions for the $\alpha_1$th and $\alpha_2$th field modes are given by:
\begin{eqnarray}
\frac{v_1\Omega_{\alpha_1}}{c_s} &=& \Omega_{\alpha_1} + \omega_{d_1}, \cr
\frac{v_2\Omega_{\alpha_2}}{c_s} &=& \Omega_{\alpha_2} + \omega_{d_2}.
\label{resonance2}
\end{eqnarray}
It is crucial to have discrete chain phonon modes, as they allow the selective excitation of certain modes that match the resonance conditions. The interaction between the detector and these phonon modes allows only the relevant modes to be excited, which is fundamental for the detector’s internal state and the chain's phonon state to accurately reflect the detector's trajectory. Additionally, there should be no chain phonon frequency $\Omega_{\alpha}$ that satisfies the conditions $v_1\Omega_{\alpha}/c_s = \Omega_{\alpha} + \omega_{d_2}$ or $v_2\Omega_{\alpha}/c_s = \Omega_{\alpha} + \omega_{d_1}$. This guarantees that the detector state with internal frequency $\omega_{d_1}$ is excited only when the detector travels at $v_1$, and the detector state with internal frequency $\omega_{d_2}$ is excited only when the detector travels at $v_2$. Applying perturbation theory under the assumption that $g^\prime _{\alpha_1}$ and $g^\prime _{\alpha_2} \ll 1$, the system density matrix at time $t$ is given by $\rho^s(t) = |\psi^s(t)\rangle \langle \psi^s(t)|$, with 
\begin{eqnarray}
|\psi^s(t)\rangle&\approx&\frac{1}{\sqrt{2}}|x_{01}+v_1t\rangle_d \otimes\left(|gg\rangle \otimes \prod_{\alpha=1}^{N-1}|0\rangle_{\alpha}- \frac{ig^\prime _{\alpha_1}t}{2} |eg\rangle \otimes \prod_{\alpha\neq \alpha_1} |0\rangle_{\alpha} |1\rangle_{\alpha_1}\right)   \cr
&&+\frac{1}{\sqrt{2}}|x_{02}+v_2t\rangle_d \otimes \left(|gg\rangle \otimes \prod_{\alpha=1}^{N-1}|0\rangle_{\alpha}- \frac{ig^\prime _{\alpha_2}t}{2} |ge\rangle \otimes \prod_{\alpha\neq \alpha_2} |0\rangle_{\alpha} |1\rangle_{\alpha_2}\right),
\label{sysdensma}
\end{eqnarray}
where $g^\prime _{\alpha_1}=-\frac{g\hbar \Omega_{\alpha_1}} {w^2c_s^2} \sqrt{\frac{2\Omega_{\alpha_1}} {\rho_c L \tilde{m}_d\omega_{d_1}}}f(\Omega_{\alpha_1} w/c_s)$, $g^\prime _{\alpha_2}=-\frac{g\hbar \Omega_{\alpha_2}} {w^2c_s^2} \sqrt{\frac{2\Omega_{\alpha_2}} {\rho_c L \tilde{m}_d\omega_{d_2}}}f(\Omega_{\alpha_2} w/c_s)$. In this case, the detector trajectories get entangled with its internal frequency excitations. 

When tracing out the chain phonon and the detector's position degrees of freedom, we are left with the reduced density matrix describing the detector's internal degree of freedom:
\begin{eqnarray}
\rho^s_d(t) \approx \frac{1}{\mathcal{N}'} \left\{|gg\rangle \langle gg|+ \frac{1}{8}  \left[(g^\prime _{\alpha_1} t)^2|eg\rangle \langle eg| + (g^\prime _{\alpha_2} t)^2 |ge\rangle \langle ge| \right] \right\},
\label{reduced1}
\end{eqnarray}
where $\mathcal{N}'=1+[(g^\prime _{\alpha_1} t)^2+(g^\prime _{\alpha_2} t)^2]/8$ normalizes the final state.

One might wonder what the detector state would be like if it travels at a fully mixed state of velocities $v_1$ and $v_2$. Predictably, the resulting density matrix is identical to the one described previously. However, in the case where the detector travels at a localized state of either velocity, denoted as situation ($l_1$) and ($l_2$), the density matrix would be as follows:
\begin{eqnarray}
\rho^{l_1}_d(t) &\approx& \frac{1}{1+(g^\prime_{\alpha_1} t)^2/4} \left\{|gg\rangle \langle gg|+ \frac{1}{4} (g^\prime_{\alpha_1} t)^2 |eg\rangle \langle eg| \right\} ,
\label{reduced2}
\end{eqnarray}
\begin{eqnarray}
\rho^{l_2}_d(t) &\approx& \frac{1}{1+(g^\prime_{\alpha_2} t)^2/4} \left\{|gg\rangle \langle gg|+ \frac{1}{4} (g^\prime_{\alpha_2} t)^2 |ge\rangle \langle ge| \right\}.
\label{reduced3}
\end{eqnarray}
Comparing Eqs.~(\ref{reduced1}),~(\ref{reduced2}) and~(\ref{reduced3}), one can observe that at time $t$,  a detector traveling at a superposition of velocities $v_1$ and $v_2$ would have a nonzero transition rate for both modes $\omega_{d_1}$ and $\omega_{d_2}$. In contrast, a detector traveling at a localized velocity $v_1$ would have a nonzero transition rate only for mode $\omega_{d_1}$, while a detector traveling at a localized velocity $v_2$ would have a nonzero transition rate only for mode $\omega_{d_2}$. Thus, the detector's internal state can serve as an observable to distinguish between a detector prepared in a superposition of trajectories and one traveling along a localized trajectory. As mentioned above, the excitation of the $\alpha_1$th chain phonon mode is accompanied by the excitation of the detector mode $\omega_{d_1}$. Similarly, the $\alpha_2$th chain phonon mode and the detector mode $\omega_{d_2}$ are produced in pairs, as evident from Eq.~(\ref{sysdensma}). Consequently, the detector’s internal state can act as a monitor of the chain phonon state, and the chain phonon state can similarly reflect the detector’s internal state and its trajectory.

\section{conclusion and discussion}\label{conclusion}

To show the emergence and further observe the quantum acoustic counterpart to the Ginzburg effect, we have developed a toy model comprising a dipole chain and an atomic detector. This model allows exploration of detector excitation and acoustic chain phonons from the vacuum. Our work reveals the entanglement between the detector's trajectories (velocities), its internal excitation levels, and the frequencies of vacuum-induced chain phonons. Under certain conditions, both the chain and the detector excitations can act as measurement devices for distinguishing the detector's trajectory, thus assisting the observation of the acoustic Ginzburg effect.

A pertinent question arises from the above discussion: Consider the detector as being in either a coherent superposition of two Gaussian wave packets or an incoherent mixture characterized by its momentum distribution. Can these two trajectory scenarios be distinguished through chain-based or detector-based measurements? Unfortunately, as briefly explained in the paragraph following Eq.~(\ref{reduced1}), such a distinction is not possible. Even simultaneous measurements of the chain phonon state and the detector's internal state fail to resolve this difference, as the entanglement between the detector's trajectory, internal excitation levels, and the chain phonons results in a loss of coherence after tracing out the trajectory degrees of freedom. This indicates the potential inability of indirect measurement schemes to distinguish these trajectories. To address this limitation, we conjecture that a direct measurement of the detector's trajectory state is necessary. For example, simultaneously measuring the detector’s trajectory and chain phonon production could lead to different excitation probabilities of the detector's internal degree of freedom, depending on whether the trajectory is coherent or incoherent. Further investigation of alternative measurement strategies or experimental setups is needed, and we leave this as an open question.


The finite dipole chain can serve as an acoustic analogue of a cavity vacuum, characterized by finite cavity length and discrete cavity modes. Building on this foundation, future research could investigate scenarios where the detector's center-of-mass travels in superpositions of accelerations, potentially exploring phenomena such as dynamical Casimir effect and the Unruh effect. This approach could also provide insights into quantum gravity within the framework of the weak equivalence principle. Additionally, considering the detector's superposition of antiparallel velocities or accelerations might lead to interesting studies on the interference effects of phonons produced from the vacuum.
\section{Acknowledgements}
The author thanks Miles Blencowe, Alexander R. H. Smith, and Flaminia Giacomini for invaluable discussions and feedback. This work was supported by the U.S. Department of Energy (DE-SC-0023103, FWP-ERW7011, DESC0024882); the Welch Foundation (A-1261); the National Science Foundation (PHY-2013771); and the Air Force Office of Scientific Research (FA9550-20-1-0366).

\onecolumngrid
\appendix
\section{\label{coordtrans} Lagrangians with Coordinate transformation}
The individual Lagrangians of the detector and chain subsystems are given by  
\begin{eqnarray}
L_\mathrm{det} &=& \frac{1}{2} m_p \dot{x}_p^2 + \frac{1}{2} m_e \dot{x}_e^2 - \frac{1}{2} k_d (x_e - x_p - a_d)^2,  
\label{ldetector1}
\end{eqnarray}  
\begin{equation}
L_\mathrm{chain} = \frac{1}{2} m_c \sum_{n=-\frac{N-1}{2}}^{\frac{N-1}{2}} \dot{x}_{c,n}^2 - \frac{1}{2} k_c \sum_{n=-\frac{N-1}{2}}^{\frac{N-3}{2}} (x_{c,n+1} - x_{c,n} - a_c)^2.  
\label{ruler1}
\end{equation}

The Coulomb interaction potential energy between the chain and detector is expressed as  
\begin{eqnarray}
V_\mathrm{int} &=& \frac{q_d q_c}{4\pi \epsilon_0} \sum_{n=-\frac{N-1}{2}}^{\frac{N-1}{2}} \left[\frac{1}{\sqrt{(w-l/2)^2+(x_p-x_{c,n})^2}} - \frac{1}{\sqrt{(w+l/2)^2+(x_p-x_{c,n})^2}} \right. \cr  
&&\left. - \frac{1}{\sqrt{(w-l/2)^2+(x_e-x_{c,n})^2}} + \frac{1}{\sqrt{(w+l/2)^2+(x_e-x_{c,n})^2}} \right] \cr  
&\approx& \frac{q_d {\mathsf{p}}_c w}{4\pi \epsilon_0} \sum_{n=-\frac{N-1}{2}}^{\frac{N-1}{2}} \left\{ \left[w^2+(x_p-x_{c,n})^2\right]^{-3/2} - \left[w^2+(x_e-x_{c,n})^2\right]^{-3/2} \right\}.
\label{coulombeq}
\end{eqnarray}  
The approximation in the last line assumes $l \ll w$, where $l$ is the separation between the chain dipole charges $q_c$ and $-q_c$. Under this condition, the potential is Taylor expanded to the first order of $l$, yielding the approximation. Here, ${\mathsf{p}}_c = q_c l$ denotes the chain’s electric dipole moment. 

We now transform to center-of-mass and relative coordinates for the chain-detector system. For the chain, the center-of-mass and relative coordinates are defined as  
\begin{equation}
x_\mathrm{cCM} = \frac{1}{N} \sum_{n=-\frac{N-1}{2}}^{\frac{N-1}{2}} x_{c,n},  
\end{equation}  
\begin{equation}
\phi_{n} = x_{c,n} - x_\mathrm{cCM} - n a_c.  
\end{equation}  
Here, $x_\mathrm{cCM}$ is the chain's center-of-mass position, and $\phi_n$ gives the displacement of the $n$-th dipole relative to its classical equilibrium position.  

For the detector subsystem, the center-of-mass and relative coordinates are given by  
\begin{eqnarray}
x_d = \frac{m_p x_p + m_e x_e}{m_p + m_e}, \quad x_{ep} = x_e - x_p,  
\end{eqnarray}  
where $x_d$ represents the center-of-mass position and $x_{ep}$ describes the relative displacement between the two detector components. The total and reduced masses of the detector are defined as $M_d = m_p + m_e$ and $\tilde{m}_d = m_p m_e / (m_p + m_e)$, respectively.  

Using these coordinates, the chain Lagrangian becomes  
\begin{equation}
L_\mathrm{chain} = \frac{1}{2} M_c \dot{x}_\mathrm{cCM}^2 + \frac{1}{2} m_c \sum_{n=-\frac{N-1}{2}}^{\frac{N-1}{2}} \dot{\phi}_{n}^2 - \frac{1}{2} k_c \sum_{n=-\frac{N-1}{2}}^{\frac{N-3}{2}} (\phi_{n+1} - \phi_{n})^2,  
\end{equation}  
where $M_c = N m_c$ is the chain’s total mass. The detector Lagrangian becomes  
\begin{eqnarray}
L_\mathrm{det} &=& \frac{1}{2} M_d \dot{x}_d^2 + \frac{1}{2} \tilde{m}_d \dot{x}_{ep}^2 - \frac{1}{2} k_d (x_{ep} - a_d)^2.  
\end{eqnarray}  
The Coulomb interaction potential energy (\ref{coulombeq}), expressed in terms of the chain dipole displacement, the detector's center-of-mass, and relative coordinates, is given by:
\begin{eqnarray}
V_\mathrm{int} &=& \frac{q_d {\mathsf{p}}_c}{4\pi \epsilon_0} \sum_{n=-\frac{N-1}{2}}^{\frac{N-1}{2}} \left\{ \left[ \left(-\frac{m_e}{m_p+m_e} x_{ep} - \phi_n - n a_c + x_d - x_\mathrm{cCM} \right)^2 + w^2 \right]^{-3/2} \right. \cr  
&&\left. - \left[ \left(\frac{m_p}{m_p+m_e} x_{ep} - \phi_n - n a_c + x_d - x_\mathrm{cCM} \right)^2 + w^2 \right]^{-3/2} \right\} .
\end{eqnarray}  
To approximate the potential energy, we consider the limit where $|x_{ep}|, |\phi_n| \ll w$. Accordingly, we use a Taylor expansion up to the second order of the potential in $x_{ep}$ and $\phi_n$ to derive the simplified form:
\begin{eqnarray}
V_\mathrm{int} \approx \frac{q_d {\mathsf{p}}_c w}{4\pi \epsilon_0} \sum_{n=-\frac{N-1}{2}}^{\frac{N-1}{2}} \left.\left[ -\frac{x_{ep}}{a_c} \frac{\partial}{\partial n} (x_n^{\prime 2} + w^2)^{-3/2} + \frac{x_{ep} \phi_n}{a_c^2} \frac{\partial^2}{\partial n^2} (x_n^{\prime 2} + w^2)^{-3/2} \right]\right|_{x_n^\prime=x_d - na_c- x_\mathrm{cCM}}.
\end{eqnarray} 
We retain the second order of the Taylor expansion to preserve the interaction between the detector's internal degree of freedom and the chain dipole displacement.


\section{\label{hamcontlim}Detector-chain Hamiltonian in the continuum limit}

In this appendix, we derive the Detector-Chain Hamiltonian in the continuum limit. To simplify the analysis, we eliminate $x_\text{cCM}$ by selecting the chain's center of mass as the observer's reference frame. Since the free detector Hamiltonian does not affect the detector-chain Hamiltonian in the continuum limit, we further simplify the model by freezing the atomic detector's internal degree of freedom (henceforth omitting the term ``detector" in this appendix), setting $|x_{ep}| = a_d$. With this, the Hamiltonian operator is defined as $\hat{H}=\hat{H}_{\mathrm{chain}}+\hat{V}_{\mathrm{int}}$, following Eqs. (\ref{chainham}) and (\ref{linterac2}):
\begin{eqnarray}
    \hat{H}&=&\sum_{n=-\frac{N-1}{2}}^{\frac{N-1}{2}} \frac{\hat{p}_{n}^2}{2m_c} + \frac{1}{2} k_c \sum_{n=-\frac{N-1}{2}}^{\frac{N-3}{2}} (\hat{\phi}_{n+1} - \hat{\phi}_{n})^2+ \frac{\hat{p}^2_d}{2 M_d} \cr
    &&+\frac{{\mathsf{p}}_d {\mathsf{p}}_c w} {4\pi \epsilon_0} \sum_{n=-\frac{N-1}{2}} ^{\frac{N-1}{2}} \left\{ -\frac{1}{a_c} \frac{\partial}{\partial n} \left[\left( na_c - \hat{x}_d \right)^2 + w^2\right]^{-3/2} + \frac{ \hat{\phi}_n} {a_c^2} \frac{\partial^2}{\partial n^2} \left[\left( na_c - \hat{x}_d \right)^2 + w^2\right]^{-3/2} \right\}.
\label{hamtoteq}
\end{eqnarray}
Hamiltonian (\ref{hamtoteq}) yields the following Hamilton's equations of motion for the atom and relative dipole coordinates, respectively:
\begin{equation}
    \frac{d \hat{x}_d}{dt}=\frac{\hat{p}_d}{M_d},
    \label{atomposeq}
\end{equation}
\begin{equation}
\frac{d \hat{p}_d}{dt}=\frac{{\mathsf{p}}_d {\mathsf{p}}_c w} {4\pi \epsilon_0} \sum_{n=-\frac{N-1}{2}} ^{\frac{N-1}{2}} \left\{ -\frac{ 1} {a_c^2} \frac{\partial^2}{\partial n^2} \left[\left( na_c - \hat{x}_d \right)^2 + w^2\right]^{-3/2} + \frac{\hat{\phi}_n} {a_c^3} \frac{\partial^3} {\partial n^3} \left[\left( na_c - \hat{x}_d \right)^2 + w^2\right]^{-3/2} \right\}  ,
\label{atommomeq}
\end{equation}
\begin{equation}
    \frac{d \hat{\phi}_n}{dt}=\frac{\hat{p}_n}{m_c},
    \label{rulerposeq}
\end{equation}
\begin{equation}
\frac{d \hat{p}_{n}}{dt}=k_c \left(\hat{\phi}_{n+1}-2 \hat{\phi}_n +\hat{\phi}_{n-1}\right)-\frac{{\mathsf{p}}_d {\mathsf{p}}_c w} {4\pi \epsilon_0} \frac{1}{a_c^2} \frac{\partial^2}{\partial n^2} \left[\left( na_c - \hat{x}_d \right)^2 + w^2\right]^{-3/2} .
\label{rulermomeq}
\end{equation}

In the limit $N\gg 1$ with $L=(N-1) a_c\gg  w$, a continuum approximation for the dipole displacements can be made: $\hat{\phi}_n(t)\rightarrow \hat{\phi(}x,t)$ and $\hat{p}_n(t)\rightarrow \hat{\pi}(x,t)$  with $x=na_c$. The sums over dipoles in the kinetic and potential energy terms of the field Lagrangian can then be approximated by the following integrals:
\begin{eqnarray}
\frac{1}{2} m_a \sum_{n=-\frac{N-1}{2}} ^{\frac{N-1}{2}} \dot{\hat{\phi}}_n(t)^2 &\to& \frac{1}{2} \frac{m_a}{a_c} \int_{-L/2}^{L/2} dx \dot{\hat{\phi}}(x,t)^2\cr
\frac{1}{2} k_c \sum_{n=-\frac{N-1}{2}} ^{\frac{N-3}{2}} \left[\hat{\phi}_{n+1}(t) - \hat{\phi}_n(t)\right]^2 &\to& \frac{1}{2} k_c a_c \int_{-L/2}^{L/2} dx \hat{\phi}^{\prime} (x,t)^2.
\end{eqnarray}
With this continuum approximation, the above Hamilton's equations (\ref{rulerposeq}), (\ref{rulermomeq}) for the dipole coordinates become:
\begin{equation}
    \frac{\partial \hat{\phi}(x,t)}{\partial t}=\frac{\hat{\pi}(x,t)} {\rho_c},
    \label{rulerconposeq}
\end{equation}
\begin{equation}
\frac{\partial \hat{\pi}(x,t)}{\partial t}=\Upsilon_c \frac{\partial^2 \hat{\phi}(x,t)}{\partial x^2}-\frac{{\mathsf{p}}_d {\mathsf{p}}_c w} {4\pi \epsilon_0 a_c}  \frac{\partial^2}{\partial x^2} \left[\left(x- \hat{x}_d \right)^2 + w^2\right]^{-3/2},
\label{rulerconmomeq}
\end{equation}
where $\rho_c=m_c/a_c$ is the chain mass density, and $\Upsilon_c=k_c a_c$ is the Young's modulus. Hamilton's equation (\ref{atommomeq}) for the atom's momentum operator becomes
\begin{equation}
\frac{d \hat{p}_d}{dt}= \frac{{\mathsf{p}}_d {\mathsf{p}}_c w} {4\pi \epsilon_0 a_c} \int_{-L/2}^{L/2} dx \left\{ -\frac{\partial^2}{\partial x^2} \left[\left(x - \hat{x}_d \right)^2 + w^2\right]^{-3/2} + \hat{\phi}(x,t) \frac{\partial^3}{\partial x^3} \left[\left(x - \hat{x}_d \right)^2 + w^2\right]^{-3/2} \right\}.
\label{atommomconeq}
\end{equation}
The Hamiltonian corresponding to Hamilton's equations (\ref{atomposeq}), (\ref{rulerconposeq}-\ref{atommomconeq}) can then be approximated as: 
\begin{eqnarray}    
   \hat{H} &=&\frac{1}{2}\int_{-L/2}^{L/2} dx \left( \frac{\hat{\pi}^2} {\rho_c} + \Upsilon_c \nabla\hat{\phi}^2 \right) + \frac{\hat{p}^2_d}{2 M_d}  \cr
    &&+ \frac{{\mathsf{p}}_d {\mathsf{p}}_c w}{4\pi \epsilon_0 a_c} \int_{-L/2} ^{L/2}dx\left\{-\frac{\partial}{\partial x} \left[(x - \hat{x}_d)^2 + w^2\right]^{-3/2}+\hat{\phi}(x,t)  \frac{\partial^2}{\partial x^2}\left[(x - \hat{x}_d)^2 + w^2\right]^{-3/2}\right\}.
     \label{hamcontinuumeq}
\end{eqnarray}
We will then restore (henceforth using the term `atomic detector')  the atom's internal, oscillator degree of freedom $x_{ep}$. This allows the detector-chain system to model the (Galilean) Unruh effect, Hawking effect, and Ginzburg radiation, where the detector's internal degree of freedom can become excited. The total, charged, detector-chain system Hamiltonian is $\hat{H}_{\mathrm{tot}}=\hat{H}_{\mathrm{det}}+\hat{H}$:
\begin{eqnarray}    
   \hat{H}_{\mathrm{tot}} &=& \frac{1}{2}\int_{-L/2}^{L/2} dx \left( \frac{\hat{\pi}^2} {\rho_c} + \Upsilon_c \nabla\hat{\phi}^2 \right) +\frac{ p_d^2} {2M_d} +\frac{p_{ep}^2} {2\tilde{m}_d} + \frac{1}{2} k_d (\hat{x}_{ep}-a_d)^2\cr
    &&+ g \hat{x}_{ep}\int_{-L/2} ^{L/2}dx\left[-\frac{\partial}{\partial x} h(x,\hat{x}_d)+\phi  \frac{\partial^2}{\partial x^2}h(x,\hat{x}_d) \right],\cr
    &&\label{continu}
\end{eqnarray}
where we have used the shorthand notation $g = {\mathsf{p}}_d {\mathsf{p}}_c w/ ({4\pi \epsilon_0 a_d a_c})$ and $h(x,\hat{x}_d) = \left[\left( x - \hat{x}_d \right)^2 + w^2\right]^{-3/2}$.

\section{\label{semiclameth} Semiclassical methods}
The Heisenberg equations for the momentum operators $\hat{p}_{ep}$, $\hat{p}_d$, and $\hat{\pi}$ can be directly derived from the Hamiltonian in Equation (\ref{continuum1}) as follows:
  \begin{equation}
\dot{\hat{p}}_{ep}=-k_d (\hat{x}_{ep}-a_d)-g \int_{-L/2} ^{L/2}dx\left[\frac{\partial}{\partial x} h(x,\hat{x}_d) +\hat{\phi} \frac{\partial^2}{\partial x^2} h(x,\hat{x}_d)\right],
\label{bdeq}
\end{equation}
 \begin{equation}
\dot{\hat{p}}_d=g \hat{x}_{ep} \int_{-L/2} ^{L/2} dx \left[\frac{\partial^2}{\partial x^2} h(x,\hat{x}_d)+ \hat{\phi} \frac{\partial^3}{\partial x^3} h(x,\hat{x}_d)\right],
\label{atommom2eq}
\end{equation}
and
 \begin{equation}
\frac{\partial \hat{\pi}} {\partial t}=\Upsilon_c \frac{\partial^2\hat{\phi}} {\partial x^2}-g  \hat{x}_{ep}  \frac{\partial^2}{\partial x^2} h(x,\hat{x}_d).
\end{equation}
The Heisenberg equations for the canonical position operators are:
\begin{equation}
 \dot{\hat{x}}_{ep}(t) ={\hat{p}_{ep}(t)} /{\tilde{m}_d},
\end{equation}
\begin{equation}
 \dot{\hat{x}}_d(t) = {\hat{p}_d(t)}/ {M_d},
\end{equation}
\begin{equation}
 \partial{\hat{\phi}}/\partial{t} = \hat{\pi}/ {\rho_c}.
\end{equation}
To apply the semiclassical method, we decompose the operators of the detector's internal degree of freedom and the chain field, along with their conjugate momenta, into their classical mean coordinates and quantum fluctuating components, as expressed in Eq.~(\ref{decomposition}). Additionally, assuming that the detector's center of mass follows a classical trajectory, we have $\hat{x}_d(t)=\bar{x}_d(t)$ and $\hat{p}_d(t)=\bar{p}_d(t)$. Substituting these decompositions into the Heisenberg equations derived above, the resulting equations are separated into two sets: (i) mean coordinate equations for the detector and the field, and (ii) equations for the quantum fluctuation terms, which depend on the mean coordinates. 

Assuming a weak detector-field coupling and that the detector's internal degree of freedom is initially in the ground state, the interaction has a negligible effect on the mean dynamics of the detector's center of mass and its internal degree of freedom. Consequently, the equations of motion for the mean coordinate solutions of the detector's internal degree of freedom reduce to:
\begin{eqnarray}
   \dot{ \bar{p}}_{ep}(t) &=& 0, \cr
    \dot{ \bar{x}}_{ep}(t) &=& 0
    \label{avephieq1}
\end{eqnarray}
with $\bar{x}_{ep}(t) = a_d$.

For the case where the detector’s center of mass is initially moving along a localized trajectory at speed $v$, the dynamics are described by:
\begin{eqnarray}
\bar{p}_d(t) &=& M_d v, \cr \bar{x}_d(t) &=& x_0+ v t.
\end{eqnarray}

The dynamical equations of the mean chain field coordinates $\bar{\phi}(x,t)$ and $\bar{\pi}(x,t)$ are
\begin{eqnarray}
 \frac{\partial{\bar{\pi}}}{\partial t} &=& \Upsilon_c \frac{\partial^2 \bar{\phi}} {\partial x^2} -g a_d  \frac{\partial^2}{\partial x^2} h[x,\bar{x}_d(t)]\cr
 \frac{\partial\bar{\phi}}{\partial t} &=& \frac{\bar{\pi}} {\rho_c}.
 \label{dynamical}
\end{eqnarray}
Assuming the chain-detector interaction starts at $t_0=0$, the dynamical equations~(\ref{dynamical}) yield the following solution, which corresponds to Eq.~(\ref{avephieq}) in the main text:
\begin{eqnarray}
    \bar{\phi}(x,t)= -g \sum_{\alpha=1}^{N-1}u_{\alpha}(x) \int_0^{t} dt' \frac{\sin\left[\Omega_{\alpha} (t-t')\right]}{\Omega_{\alpha}} \int_{-L/2}^{L/2} dx' \frac{\partial^2}{\partial x^{\prime 2}} h[x',\bar{x}_d(t')] u_{\alpha}(x'),
    \label{phisolneq}
\end{eqnarray}
where $\Omega_\alpha$ are the $\alpha$-th normal mode frequencies of the chain, and $u_{\alpha}(x)$ are the $\alpha$-th orthonormal mode eigenfunctions of the chain:
\begin{eqnarray}
    \Omega_\alpha&=&2\sqrt{\frac{k_c}{m_c}} \sin \left(\frac{\alpha\pi a_c}{2L}\right) ,\, \alpha=1,2,\dots,N-1,\cr
    u_{\alpha}(x)&=&\sqrt{\frac{2}{L}}\cos\left[\frac{\alpha \pi}{L} \left(x+\frac{L}{2}\right)\right],\, \alpha=1,2,\dots,N-1,
\end{eqnarray}
with $L=(N-1) a_c$ is the classical equilibrium chain length, and the subscript $\alpha$ denotes $\alpha$-th phonon mode.

We then expand the Heisenberg equations up to the second order in the quantum fluctuation terms. The dynamical equations for the quantum position operators are given by

$\delta\dot{\hat{x}}_{ep}(t) ={\delta \hat{p}_{ep}(t)} /{\tilde{m}_d}$ and $\partial\delta \hat{\phi}(x,t) /\partial t = {\delta \hat{\pi}(x,t)}/ {\rho_c}$. The fact that $h(x,\bar{x}_d) = \left\{\left[ x-\bar{x}_d(t)\right]^2 + w^2\right\}^{-3/2}$ is an even function with respect to $\bar{x}_d$ for $L\gg w,\, |\bar{x}_d|$ allows us to omit certain terms from the dynamical equations for the corresponding quantum momentum operators. This results in the following simplification:
\begin{eqnarray}
    \delta\dot{\hat{p}}_{ep} &=& -k_d\delta \hat{x}_{ep}- g \int_{-L/2}^{L/2} dx  \delta \hat{\phi} \frac{\partial^2}{\partial x^2} h(x,\bar{x}_d),
    \label{flucdyna1}
\end{eqnarray}   
\begin{eqnarray}    
    \frac{\partial\delta \hat{\pi}} {\partial t} &=& \Upsilon_c \frac{\partial^2 \delta \hat{\phi}}{\partial x^2} - g \delta \hat{x}_{ep} \frac{\partial^2}{\partial x^2} h(x,\bar{x}_d).
    \label{flucdyna2}
\end{eqnarray}
The Hamiltonian for the quantum fluctuations, derived from the Heisenberg equations for $\delta \hat{x}_{ep}(t)$ and $\delta\hat{\phi}(x,t)$, along with Eqs.~\eqref{flucdyna1} and \eqref{flucdyna2}, is expressed as a sum of quadratic terms:
\begin{eqnarray}    
       \hat{H}_\mathrm{Q} = \frac{1}{2}\int_{-L/2}^{L/2} dx \left[\frac{\delta\hat{\pi}^2} {\rho_c} + \Upsilon_c (\nabla\delta \hat{\phi})^2 \right] +\frac{\delta \hat{p}_{ep}^2}{2\tilde{m}_d} + \frac{1}{2} k_d \delta \hat{x}_{ep}^2+g \int_{-L/2}^{L/2} dx \delta\hat{\phi} \delta \hat{x}_{ep}\frac{\partial^2}{\partial x^2} h(x,\bar{x}_d).
\end{eqnarray}

\section{\label{localizeatom} Chain response to traveling detector}
The average displacement of a given dipole from its reference coordinate location $x$ at time $t$ is described by $\bar{\phi}(x,t)$. Here we ignore the detector's internal degree of freedom for simplicity, assume that the dipole chain is initially in its equilibrium state, and the detector locates at $x_0$ at $t=0$. 

Equation (\ref{avephieq}) that describes $\bar{\phi}(x,t)$ is written as
\begin{equation}
    \bar{\phi}(x,t)=-g \sum_{\alpha=1}^{N-1} u_{\alpha}(x)\int_0^{t} dt' \frac{\sin \left[\Omega_{\alpha} (t-t')\right]}{\Omega_{\alpha}} \int_{-L/2}^{L/2} dx' \frac{\partial^2}{\partial x^{\prime 2}} h[x',\bar{x}_d(t')]u_{\alpha}(x'),
    \label{avephieq2}
\end{equation}

Carrying out the time integral, and evaluating the spatial integral by parts with the detector assumed to be well away from the chain edges (i.e., $|\bar{x}_d(t)|\ll L$), $\bar{\phi}(x,t)$ can be solved analytically:
\begin{eqnarray}
   \bar{\phi}(x,t) &\approx& -\sum_{\alpha=1}^{N-1}\frac{2g a_d} {\rho_c w^2} \sqrt{\frac{L}{2}} u_{\alpha}(x)f(\Omega_{\alpha} w/c_s) \left\{\frac{1}{Lc_s(c_s-v)} \cos\left[\frac{\Omega_\alpha}{c_s} \left(x_0+ vt_0 + c_st+\frac{L}{2}\right)\right]\right.\cr
    &&-\frac{2}{L(c_s^2-v^2)} \cos\left[\frac{\Omega_\alpha}{c_s} \left(x_0+ vt+\frac{L}{2}\right)\right]\cr
    &&\left.+\frac{1}{Lc_s(c_s+v)} \cos\left[\frac{\Omega_\alpha}{c_s} \left(x_0+vt_0- c_st+\frac{L}{2}\right)\right]\right\},
\label{analyticalfield}
\end{eqnarray}
where $f(y)=y K_1(y)$ is a cut-off function around $y\sim 1$, with $f(0)=1$, and $K_1(y)$ is the modified Bessel function of the second kind, given by:
\begin{equation}
 K_1(y)=y \int_0^{\infty} dx (x^2+y^2)^{-3/2} \cos x.
\end{equation}
The cut-off function $f(\Omega_{\alpha} w/c_s)$ suppresses phonon modes with wavelength smaller than $w$.

Next, we focus on the second line of the analytical expression~(\ref{analyticalfield}) of $\bar{\phi}(x,t)$, which is:  
\begin{eqnarray}
-\frac{2}{L(c_s^2-v^2)} \cos\left[\frac{\Omega_\alpha}{c_s} \left(x_0+ vt+\frac{L}{2}\right)\right].
\end{eqnarray}
We rewrite this term with the prefactor from Eq.~(\ref{analyticalfield}), and denote the sum of these terms as $\bar{\phi}_2(x,t)$. The expanded solution for $\bar{\phi}_2(x,t)$ is given by:
\begin{eqnarray}
    \bar{\phi}_2(x,t) = \frac{2g a_d} {\rho_c L w^2} \frac{2} {c_s^2-v^2} \sum_{\alpha=1}^{N-1} f(\Omega_{\alpha} w/c_s) \cos\left[\frac{\Omega_\alpha}{c_s} \left(x+\frac{L}{2}\right)\right] \cos\left[\frac{\Omega_\alpha}{c_s} \left( \bar{x}_d(t)+\frac{L}{2}\right)\right].
\label{phimain}
\end{eqnarray}
We next consider the chain mode continuum limit, assumed valid for large chain length $L$. 
By defining $\tilde{\Omega}=\Omega_{\alpha}w/c_s$, Eq.~(\ref{phimain}) becomes approximately
\begin{eqnarray}
    \bar{\phi}_2(x,t) &=&\frac{g a_d} {\rho_c \pi w^3} \frac{2} {c_s^2-v^2} \int_0^{\infty} d\tilde{\Omega} \tilde{\Omega}^2 \cos\left[\frac{\tilde{\Omega}}{w} \left(x+\frac{L}{2}\right)\right] \cos\left[\frac{\tilde{\Omega}}{w} \left( \bar{x}_d(t)+\frac{L}{2}\right)\right]\cr
    &&\times \int_{-\infty}^{\infty}dx^{\prime}\left(x^{\prime 2}+\tilde{\Omega}^2\right)^{-\frac{3}{2}}\cos x^{\prime}.
\label{phicontinu}
\end{eqnarray}
Transforming to `polar' coordinates defined as $\tilde{\Omega}=r\cos\theta$ and $x^{\prime}=r\sin\theta$, Eq. (\ref{phicontinu}) becomes
\begin{eqnarray}
    \bar{\phi}_2(x,t) &=& \frac{g a_d} {\rho_c \pi w^3} \frac{1} {c_s^2-v^2}\int_0^{2\pi} d\theta\cos^2\theta \int_0^{\infty} D \cos\left[\frac{r\cos\theta}{w} \left(x+\frac{L}{2}\right)\right] \cr
    &&\times \cos\left[\frac{r\cos\theta}{w} \left( \bar{x}_d(t)+\frac{L}{2}\right)\right] \cos\left(r\sin\theta\right).
\label{phipolar}
\end{eqnarray}
The polar integrals can be carried out analytically and we arrive at the following expression for $\bar{\phi}_2(x,t)$:
\begin{eqnarray}
    \bar{\phi}_2(x,t) &=& \frac{g a_d} {\rho_c w^3} \frac{1} {c_s^2-v^2} \left\{\left[ 1+\left( \frac{x+\bar{x}_d(t)+L} {w}\right)^2 \right]^{-\frac{3}{2}} +\left[ 1+\left( \frac{x-\bar{x}_d(t)} {w}\right)^2 \right]^{-\frac{3}{2}}\right\}.
\label{phifinal0}
\end{eqnarray}
For $w\ll L$, the first term is negligible compared to the second term, and  we arrive at the following concise approximation to the mean field chain response to the detector: 
\begin{eqnarray}
    \bar{\phi}_2(x,t) &=& \frac{g a_d} {\rho_c w^3 } \frac{1} {c_s^2-v^2} \left[ 1+\left( \frac{x-\bar{x}_d(t)} {w}\right)^2 \right]^{-\frac{3}{2}}\cr
    &=&\frac{g a_d} {\rho_c} \frac{1} {c_s^2-v^2} \left\{\left[ x-\bar{x}_d(t)\right]^2 + w^2\right\}^{-3/2}.
\label{phifinal2}
\end{eqnarray}


We then apply similar procedure to the other two terms of Eq.~(\ref{analyticalfield}), and include them in $\bar{\phi}(x,t)$ to obtain
\begin{eqnarray}
     \bar{\phi}(x,t)=\frac{g a_d} {\rho_c} \left[\frac{h(x,\bar{x}_d)} {c_s^2-v^2} -\frac{h(x,x_0+c_st)}{2c_s (c_s-v)}  -\frac{h(x,x_0-c_st)}{2c_s (c_s+v)} \right].
\label{phifinal3}
\end{eqnarray}

\section{\label{systate} General Approach to Chain-Detector Dynamics}

\subsection{\label{subsec:single} Detector with a single internal degree of freedom}
Consider a detector prepared in a superposition of two localized positions in space, characterized by a relative phase $\phi$ and a parameter $\theta$ that defines the weight of each localized trajectory. The initial state of the chain-detector system is given by:
\begin{align}
|\psi^s(0)\rangle= \int_{-L/2}^{L/2}dx \left[ \cos\theta \delta\left(\hat{x}-x_{01}\right) + e^{i\phi}\sin\theta \delta\left(\hat{x}-x_{02}\right)\right] |x\rangle_d \otimes|g\rangle \otimes \prod_{\alpha=1}^{N-1} |0\rangle_{\alpha},
\label{inistate3}
\end{align}
where $\theta \in [0, \frac{\pi}{2})$, and $\phi \in [0, \pi)$. Here, $|x\rangle_d$ represents the position state of the detector. Under the free-particle time evolution operator, it evolves as $\exp(-{i\hat{p}_d^2t}/{2M_d}) |x\rangle_d= |x+{p_d t} /{M_d}\rangle_d$. Under the resonance conditions
\begin{eqnarray}  
\frac{v_1\Omega_{\alpha_1}}{c_s}&=&\Omega_{\alpha_1}+\omega_d,\cr
\frac{v_2\Omega_{\alpha_2}}{c_s}&=&\Omega_{\alpha_2}+\omega_d
\end{eqnarray}  
for the $\alpha_1$th and $\alpha_2$th field modes, and starting from the initial state given in Eq.~(\ref{inistate3}), the time-evolved state $|\psi^s(t)\rangle$ at time $t$, obtained using perturbation theory, is given by:
\begin{eqnarray}
|\psi^s(t)\rangle&=& \exp{\left(-i\int_0^t dt' \hat{H}_\mathrm{Q}\right)} |\psi^s(0)\rangle\cr
&\approx& \cos\theta \times \exp{\left[- i\hat{H}_0(\hat{p}_{d_1}) t\right]} \exp{\left(- i\hat{H}_\mathrm{I,\alpha_1}^\mathrm{int} t\right)} |x_{01}\rangle_d \otimes |g\rangle \otimes \prod_{\alpha=1}^{N-1} |0\rangle_{\alpha} \cr
&&+  e^{i\phi}\sin\theta \times\exp{\left[- i\hat{H}_0(\hat{p}_{d_2}) t\right]} \exp{\left(- i\hat{H}_\mathrm{I,\alpha_2}^\mathrm{int}t \right)} |x_{02}\rangle_d \otimes |g\rangle \otimes \prod_{\alpha=1}^{N-1} |0\rangle_{\alpha},
\end{eqnarray}
where the free Hamiltonian $\hat{H}_0$ is given by Eq.~(\ref{freehami3}). The interaction Hamiltonians $\hat{H}_\mathrm{I,\alpha_{1(2)}}^\mathrm{int}$ have the form:
\begin{equation}    
    \hat{H}_\mathrm{I,\alpha_{1(2)}}^\mathrm{int} =\frac{g_{\alpha_{1(2)}}}{2} (\hat{a}_{\alpha_{1(2)}}\hat{b} + \hat{a}_{\alpha_{1(2)}}^{\dag}  \hat{b}^{\dag}).
\end{equation}

The solution for $|\psi^s(t)\rangle$ is then obtained as:
\begin{eqnarray}
|\psi^s(t)\rangle&\approx& \cos\theta |x_{01}+v_1t\rangle_d \otimes \left( |g\rangle\otimes \prod_{\alpha=1}^{N-1}|0\rangle_{\alpha} -\frac{ig_{\alpha_1}t }{2} |e\rangle \otimes\prod_{\alpha\neq\alpha_1} |0\rangle_{\alpha} |1\rangle_{\alpha_1} \right)\cr
&&+ e^{i\phi}\sin\theta |x_{02}+ v_2t\rangle_d \otimes \left(|g\rangle \otimes \prod_{\alpha=1}^{N-1} |0\rangle_{\alpha} -\frac{ig_{\alpha_2}t }{2} |e\rangle \otimes \prod_{\alpha\neq\alpha_2} |0\rangle_{\alpha} |1\rangle_{\alpha_2} \right).
\label{solution}
\end{eqnarray}
Tracing out the chain phonon modes except for $\alpha=\alpha_1, \alpha_2$, the reduced state for the phonon modes $\alpha_1$ and $\alpha_2$ is given by
\begin{eqnarray}
\rho_c^s(t)\approx \frac{1}{\mathcal{N}} \left\{|0\rangle_{\alpha_1} \langle 0|\otimes |0\rangle_{\alpha_2} \langle 0|+ \frac{1}{4} \left[ (g_{\alpha_1} t \cos\theta )^2|1\rangle_{\alpha_1} \langle 1| \otimes |0\rangle_{\alpha_2} \langle 0|+ (g_{\alpha_2} t\sin\theta)^2|0\rangle_{\alpha_1} \langle 0|\otimes |1\rangle_{\alpha_2} \langle 1|\right] \right\},
\label{redgeneral}
\end{eqnarray}
where $\mathcal{N}=1+[(g_{\alpha_1} t \cos\theta )^2+(g _{\alpha_2} t \sin\theta)^2]/4$. When $\theta=\pi/4$, the two trajectories have equal weight, and Eqs.~(\ref{superpos}, \ref{redsuperpos}) in the main text is recovered from Eqs.~(\ref{solution}, \ref{redgeneral}). In contrast, when the detector travels along a localized trajectory with velocity $v_1$, which corresponds to the case where $\theta=0$, the reduced state for the phonon modes $\alpha_1$ and $\alpha_2$ would be:
\begin{eqnarray}
\rho_c^l(t) \approx \frac{1}{1+(g_{\alpha_1} t)^2/4} \left[|0\rangle_{\alpha_1} \langle 0|+ \frac{1}{4} (g_{\alpha_1} t)^2|1\rangle_{\alpha_1} \langle 1|\right]\otimes |0\rangle_{\alpha_2} \langle 0|.
\end{eqnarray}

Tracing out the detector's position state and all the chain phonon modes from Eq.~(\ref{solution}), we obtain the reduced density matrix describing the detector's internal degree of freedom:
\begin{eqnarray}
\rho_d^s(t) \approx \frac{1}{\mathcal{N}} \left\{|g\rangle \langle g|+ \frac{1}{4}  \left[(g _{\alpha_1} t\cos\theta)^2 + (g _{\alpha_2} t\sin\theta)^2 \right] |e\rangle \langle e|\right\}.
\label{detdens1}
\end{eqnarray}
In the case where the detector travels at a localized trajectory with velocity $v_1$, the density matrix of the detector's reduced state would be 
\begin{eqnarray}
\rho_d^l(t) \approx \frac{1}{1+(g_{\alpha_1} t)^2/4} \left[|g\rangle \langle g|+ \frac{1}{4} (g _{\alpha_1} t)^2 |e\rangle \langle e|\right].
\label{detdens2}
\end{eqnarray}

\subsection{\label{subsec:two} Detector with two different internal degrees of freedom}
We next focus on the case when the detector possesses two internal excitation frequencies, $\omega_{d_1}$ and $\omega_{d_2}$. Assuming the detector travels in a superposition state with velocities $v_1$ and $v_2$, the initial state of the chain-detector system is modified as follows:
\begin{eqnarray}
|\psi^s(0)\rangle= \int_{-L/2}^{L/2}dx \left[\cos\theta \delta(\hat{x}-x_{01})+ e^{i\phi}\sin\theta \delta(\hat{x}-x_{02}) \right] |x\rangle_d \otimes|gg\rangle \otimes \prod_{\alpha=1}^{N-1} |0\rangle_{\alpha},
\end{eqnarray}
where $|gg\rangle$ denotes the ground state of the detector's two internal excitation levels, $\omega_{d_1}$ and $\omega_{d_2}$.

The resonance conditions for the detector's two internal excitation levels, $\omega_{d_1}$ and $\omega_{d_2}$, and the $\alpha_1$th and $\alpha_2$th field modes are given by:
\begin{eqnarray}
\frac{v_1\Omega_{\alpha_1}}{c_s} &=& \Omega_{\alpha_1} + \omega_{d_1}, \cr
\frac{v_2\Omega_{\alpha_2}}{c_s} &=& \Omega_{\alpha_2} + \omega_{d_2}.
\end{eqnarray}
Applying perturbation theory under the assumption that $g^\prime _{\alpha_1}$ and $g^\prime _{\alpha_2} \ll 1$, the system density matrix at time $t$ is given by $\rho^s(t) = |\psi^s(t)\rangle \langle \psi^s(t)|$, with 
\begin{eqnarray}
|\psi^s(t)\rangle&=& \exp{\left(-i\int_0^t dt' \hat{H}_\mathrm{Q}\right)} |\psi^s(0)\rangle\cr
&\approx&\cos\theta \times\exp{\left[- i\hat{H}_0(\hat{p}_{d_1}) t\right]} \exp{\left(- i\hat{H}_\mathrm{I,\alpha_1}^\mathrm{int} t\right)}  |x_{01}\rangle_d \otimes |gg\rangle \otimes \prod_{\alpha=1}^{N-1} |0\rangle_{\alpha} \cr
&&+ e^{i\phi}\sin\theta \times\exp{\left[- i\hat{H}_0(\hat{p}_{d_2}) t\right]} \exp{\left(- i\hat{H}_\mathrm{I,\alpha_2}^\mathrm{int}t \right)}  |x_{02}\rangle_d \otimes |gg\rangle \otimes \prod_{\alpha=1}^{N-1} |0\rangle_{\alpha} .
\end{eqnarray}
In this case, the free Hamiltonian $\hat{H}_0$ is given by 
\begin{equation}    
 \hat{H}_0 =\frac{\hat{p}_d^2} {2M_d}+ \sum_{\alpha=1}^{N-1} \hbar\Omega_{\alpha} \hat{a}_{\alpha}^{\dag}\hat{a}_{\alpha} + \sum_{i=1,2} \hbar\omega_{d_i} \hat{b}_i^{\dag} \hat{b}_i,
\end{equation}
while the interaction Hamiltonians $\hat{H}_\mathrm{I,\alpha_{1(2)}}^\mathrm{int}$ take the form:
\begin{equation}    
    \hat{H}_\mathrm{I,\alpha_{1(2)}}^\mathrm{int} =\frac{g'_{\alpha_{1(2)}}}{2} (\hat{a}_{\alpha_{1(2)}}\hat{b}_{1(2)} + \hat{a}_{\alpha_{1(2)}}^{\dag}  \hat{b}_{1(2)}^{\dag}).
\end{equation}
Using these Hamiltonians, the approximate time-evolved state is expressed as
\begin{eqnarray}
|\psi^s(t)\rangle&\approx& \cos\theta |x_{01}+v_1t\rangle_d \otimes \left( |gg\rangle\otimes \prod_{\alpha=1}^{N-1}|0\rangle_{\alpha} -\frac{ig'_{\alpha_1}t }{2} |eg\rangle \otimes\prod_{\alpha\neq\alpha_1} |0\rangle_{\alpha} |1\rangle_{\alpha_1} \right)\cr
&&+ e^{i\phi}\sin\theta |x_{02}+ v_2t\rangle_d \otimes \left(|gg\rangle \otimes \prod_{\alpha=1}^{N-1} |0\rangle_{\alpha} -\frac{ig'_{\alpha_2}t }{2} |ge\rangle \otimes \prod_{\alpha\neq\alpha_2} |0\rangle_{\alpha} |1\rangle_{\alpha_2} \right).
\end{eqnarray}
When tracing out the chain phonon and the detector's position degrees of freedom, we are left with the reduced density matrix describing the detector's internal degree of freedom:
\begin{eqnarray}
\rho_d^s(t) \approx \frac{1}{\mathcal{N}'} \left\{|gg\rangle \langle gg|+ \frac{1}{4}  \left[(g' _{\alpha_1} t\cos\theta)^2 |eg\rangle \langle eg| + (g' _{\alpha_2} t\sin\theta)^2 |ge\rangle \langle ge| \right] \right\},
\end{eqnarray}
where $\mathcal{N}'=1+[(g^\prime _{\alpha_1} t\cos\theta)^2+(g^\prime _{\alpha_2} t\sin\theta)^2]/4$ is the normalization factor for the final state. Eqs.~(\ref{reduced1}), (\ref{reduced2}), and (\ref{reduced3}) in the main text correspond to the cases $\theta=\pi/4,0,\pi/2$, respectively.

\bibliography{bibliocherenkov}  
\end{document}